\documentclass[11pt,twoside]{article}
\usepackage{epic}
\usepackage{eepic,atmp,amssymb,amsmath,amsthm,amscd,epsfig}

\begin{document}
\newcommand{\be}{\begin{equation}} \newcommand{\ee}{\end{equation}}
\newcommand{\ba}{\begin{eqnarray}} \newcommand{\ea}{\end{eqnarray}}
\newcommand{\no}{\nonumber}
\newcommand{\Eqn}[1]{&\hspace{-0.5em}#1\hspace{-0.5em}&}
\newcommand{\varth}{\vartheta}
\newcommand{\aA}{\widehat{A}}
\newcommand{\aD}{\widehat{D}}
\newcommand{\hD}{\,\widehat{\!D}}
\newcommand{\aE}{\widehat{E}}
\newcommand{\hE}{\,\widehat{\!E}}
\newcommand{\tu}{\widetilde{u}}
\newcommand{\tx}{\widetilde{x}}
\newcommand{\ty}{\widetilde{y}}
\newcommand{\ttau}{\widetilde{\tau}}
\newcommand{\dprod}{\prod\limits}
\newcommand{\dsum}{\sum\limits}
\renewcommand{\dfrac}[2]{{\displaystyle\frac{#1}{#2}}}
\renewcommand{\tfrac}[2]{{\textstyle\frac{#1}{#2}}}
\newcommand{\chEei}[8]{\chi[{}_{#1}^{}{}_{#3}^{}{}_{#4}^{#2}{}_{#5}^{}
                            {}_{#6}^{}{}_{#7}^{}{}_{#8}^{}]}
\newcommand{\chEse}[7]{\chi[{}_{#1}^{}{}_{#3}^{}{}_{#4}^{#2}{}_{#5}^{}
                            {}_{#6}^{}{}_{#7}^{}]}
\newcommand{\chEsi}[6]{\chi[{}_{#1}^{}{}_{#3}^{}{}_{#4}^{#2}{}_{#5}^{}
                            {}_{#6}^{}]}
\newcommand{\chEfi}[5]{\chi[{}_{#1}^{}{}_{#3}^{}{}_{#4}^{#2}{}_{#5}^{}]}
\newcommand{\chAfi}[5]{\chi_{\mbox{\tiny$[$}}{}_{#1}^{}{}_{#2}^{}
                       {}_{#3}^{}{}_{#4}^{}{}_{#5}^{}{}_{\mbox{\tiny$]$}}}
\newcommand{\chDsi}[6]{\chi[{}_{#2}^{}{}_{#3}^{#1}{}_{#4}^{}{}_{#5}^{}
                            {}_{#6}^{}]}

%%% redefine section %%%
\renewcommand{\section}[1]{\subsection{\mathversion{bold}#1}}
\renewcommand{\thesubsection}{\arabic{subsection}.\hspace{-.5em}}
%%% define minisection %%%
\newcommand{\minisection}[1]{\subsubsection{\mathversion{bold}#1}}
\renewcommand{\thesubsubsection}
  {\thesubsection\hspace{.5em}\arabic{subsubsection}.\hspace{-.5em}}
%%% define equation number %%%
\makeatletter
\@addtoreset{equation}{subsection}
\renewcommand{\theequation}{\arabic{subsection}.\arabic{equation}}
\makeatother
%%%%%%%%%%%%%%%%%%%%%%%%%%%%%%%%%%%%%%%%%%%%%%%%%%%%%%%%%%%%%%%%%%%%%%%%%

%%% Title page %%%%%

\copyrightnotice{2003}{7}{421}{457}	
\setcounter{page}{421}

\title{ Seiberg--Witten Curve for $ E$-String Theory Revisited}
\author{Tohru Eguchi and Kazuhiro Sakai}
\address{Department of Physics, Faculty of Science,\\ University of
  Tokyo, Tokyo 113-0033, Japan}
\url{hep-th/0211213}
\barefootnote{UT-02-46}
\pagestyle{myheadings}
\markboth{\it Seiberg--Witten Curve for $E$-String Theory Revisited}{\it T. Eguchi and
  K. Sakai}
%E-mail: {\tt eguchi,sakai@hep-th.phys.s.u-tokyo.ac.jp}

\begin{abstract}
We discuss various properties of the Seiberg--Witten curve for the $E$-string
theory which we have obtained recently in hep-th/0203025.
Seiberg--Witten curve for the $E$-string describes the low-energy dynamics of a
six-dimensional $(1,0)$ SUSY theory when compactified on ${\bf R}^4\times T^2$.
It has a manifest affine $E_8$ global symmetry with modulus $\tau$ and
$E_8$ Wilson line parameters $\{m_i\},i=1,2,\ldots,8$ which are
associated with the geometry of
the rational elliptic surface.
When the radii $R_5,R_6$ of the torus $T^2$ degenerate $R_5,R_6\rightarrow 0$, 
$E$-string curve is reduced 
to the known Seiberg--Witten curves of
four- and five-dimensional gauge theories. 
 
In this paper we first study the geometry of rational elliptic surface
and identify the geometrical significance of the Wilson line parameters.
By fine tuning these parameters we also study degenerations of our curve
corresponding to various unbroken symmetry groups.
We also find a new way of reduction to four-dimensional theories
without taking a degenerate limit
of $T^2$ so that the $SL(2,{\bf Z})$ symmetry is left intact.
By setting some of the Wilson line parameters to special values 
we obtain the four-dimensional $SU(2)$ Seiberg--Witten theory with 
4 flavors and also a curve by Donagi and Witten describing the dynamics of
a perturbed ${\cal N}=4$ theory.
\end{abstract}

\section{Introduction}

In our previous paper \cite{ES}
we have constructed a Seiberg--Witten curve for
the $E$-string (or $E_8$ non-critical string) theory
\cite{GH,SW1,Ga1,KMV,Ga2,DKV,GMS,LMW,MNW1,MNW2,MNVW,Moh,Iq}.
Seiberg--Witten curve for 
$E$-strings describes the low-energy 
dynamics of a $(1,0)$ SUSY theory in six dimensions partially 
compactified on ${\bf R}^4\times T^2$.
$E$-string is essentially one half of the 
$E_8\times E_8$ heterotic string and
describes the degrees of freedom which appear on the M5-brane in the 
M-theory description of small $E_8$ instanton singularities.

Our curve has a manifest affine $E_8$ global symmetry
with the modulus $\tau$ of $T^2$ and possesses
$E_8$ Wilson line parameters $\{m_i\},i=1,2,\ldots,8$.
When the radii of the torus $T^2$ degenerate $R_5,R_6\rightarrow 0$,
the curve reduces to the Seiberg--Witten curves 
of four- and five-dimensional gauge theories \cite{SW2,MN1,MN2,NTY,MNW1}
and $\{m_i\}$ are identified as the mass
parameters of the matter hypermultiplets.

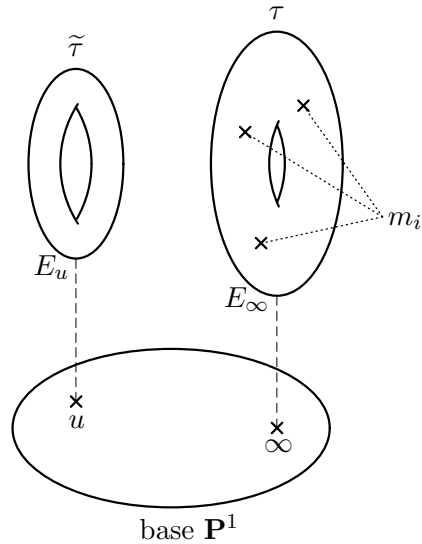
\begin{figure}[h]
\begin{center}
\unitlength=2pt
\begin{picture}(80,100)
\dottedline{1}(54,76)(80,60)
\dottedline{1}(65,81)(80,60)
\dottedline{1}(57,55)(80,60)
\dashline[50]{2}(22,25)(22,52)
\dashline[50]{2}(60,20)(60,45)
\put(81,58.5){$m_i$}
\thicklines
\put(40,20){\ellipse{60}{30}}
\put(21,24){\line(1,1){2}}
\put(21,26){\line(1,-1){2}}
\put(59,19){\line(1,1){2}}
\put(59,21){\line(1,-1){2}}
\put(20.6,20){$u$}
\put(57.6,15.5){$\infty$}
\put(34,-1){base ${\bf P}^1$}
\put(22,70){\ellipse{18}{36}}
\put(4,70){\arc{42}{0}{0.52}}
\put(4,70){\arc{42}{5.76}{6.28}}
\put(40,70){\arc{42}{2.57}{3.71}}
\put(14,49){$E_u$}
\put(20.5,90.5){$\ttau$}
\put(60,70){\ellipse{25}{50}}
\put(42.5,70){\arc{38}{0}{0.39}}
\put(42.5,70){\arc{38}{5.89}{6.28}}
\put(77.5,70){\arc{38}{2.70}{3.58}}
\put(50,43){$E_\infty$}
\put(58.5,97.5){$\tau$}
\put(53,75){\line(1,1){2}}
\put(53,77){\line(1,-1){2}}
\put(64,80){\line(1,1){2}}
\put(64,82){\line(1,-1){2}}
\put(56,54){\line(1,1){2}}
\put(56,56){\line(1,-1){2}}
\end{picture}
\caption{Elliptic fibration of rational elliptic surface \label{EF}}
\end{center}
\end{figure}

It is known for some time that the $E$-string
is associated with the geometry of the
rational elliptic surface. In fact our curve has the form 
\be
{\cal C}: \hskip3mm y^2=4x^3-f(u;\tau,m_i)x-g(u;\tau,m_i)
\label{res}\end{equation}
where the function $f,\,g$ are polynomials
of degree 4,\,6 in $u$ which denotes the 
coordinate of ${\bf P}^1$. (\ref{res}) describes the structure of an elliptic
fibration over the base ${\bf P}^1$, i.e. the rational elliptic surface.
The parameter $\tau$ is identified as the modulus of the elliptic fiber
$E_{\infty}$ at $u=\infty$. Modulus $\tilde{\tau}$ of the fiber  
at finite $u$ depends on $u$ and $\tau,m_i$ and is identified as the inverse 
gauge coupling constant in the usual manner. See Figure \ref{EF}.

Rational elliptic surface
may be constructed by blowing up 9 points of ${\bf P}^2$ and is also 
called as the (almost) del Pezzo surface ${\cal B}_9$. Its homology two-cycles
form a lattice $\Gamma^{9,1}=\Gamma_8\oplus \Gamma^{1,1}$ where $\Gamma_8$
denotes the $E_8$ root lattice and $\Gamma^{1,1}$ is a Lorentzian lattice.
Thus $b_2^{+}=1,b_2^{-}=9$.
It is well-known that the rational elliptic surfaces
appear in many ways like one half of 
the $K_3$ surface, so that they are sometimes called as the ${1\over 2}K_3$. 
In fact when M5-brane of M-theory is wrapped over $K_3$ surface, one obtains a 
heterotic string while when M5 is wrapped on ${1\over 2}K_3$,
one obtains an $E$-string \cite{MNVW}.

Using a chain of string dualities it has been pointed out
that the BPS states of $E$-string
are in one-to-one correspondence with the holomorphic curves
in ${1\over 2}K_3$ and
also the instantons of ${\cal N}=4$ gauge theory on 
${1\over 2}K_3$ \cite{MNVW}.
Let us denote the number of BPS states of $E$-string
with winding number $n$ and momentum $k$
as $N^{\mbox{\scriptsize BPS}}_{n,k}$ when it is compactified on a circle. 
Using the duality between F-theory and M-theory it is possible to 
show that $N^{\mbox{\scriptsize BPS}}_{n,k}$ 
agrees with the number $N^{\mbox{\scriptsize curve}}_{n,k}$
of holomorphic curves of ${1\over 2}K_3$ 
in the class $n[\Sigma]+k[E]$ where $\Sigma$ denotes the base ${\bf P}^1$
and $E$ the elliptic fiber. Furthermore by considering the 
M5-brane wrapped around ${1\over 2}K_3\times T^2$ it was shown that
$N^{\mbox{\scriptsize BPS}}_{n,k}$ also agrees with the
$k$ instanton amplitude $N^{\mbox{\scriptsize inst}}_{n,k}$ of
the ${\cal N}=4$ $U(n)$ gauge theory on 
${1\over 2}K_3$.
Thus we have
\ba
N^{\mbox{\scriptsize BPS}}_{n,k}
=N^{\mbox{\scriptsize curve}}_{n,k}
=N^{\mbox{\scriptsize inst}}_{n,k}\equiv N_{n,k}.
\ea
Prepotential is then defined as usual \cite{CKYZ} by
\be
{\cal F}(\phi,\tau)={\cal F}_{\mbox{\scriptsize classical}}
-\frac{1}{(2\pi i)^3}\sum_{n=0}^\infty\sum_{k=0}^\infty
N_{n,k}\mathrm{Li}_3(e^{2\pi i n\phi+2\pi i k\tau})
\end{equation}
where $\mathrm{Li}_3(x)$ is the tri-logarithm function
$\mathrm{Li}_3(x)=\sum_{m=1}^\infty(x^m/m^3)$
and $\phi$ denotes the size of the base ${\bf P}^1$.
Due to the global $E_8$ symmetry of the theory BPS states etc. of $E$-string 
fall into $E_8$ Weyl orbits and thus $N_{n,k}$ may be expanded as
\be
N_{n,k}=\sum_{\cal O}\mbox{dim}({\cal O})N^{\cal O}_{n,k}
\end{equation}
where $\mbox{dim}({\cal O})$ denotes the dimension of
the Weyl orbit $\cal O$. When we introduce the
$E_8$ Wilson line parameters $m_i,i=1,2,\ldots,8$,
the prepotential is modified as
\be\label{prepoa}
{\cal F}(\phi,\tau,\vec{m})={\cal F}_{\mbox{\scriptsize classical}}
-\frac{1}{(2\pi i)^3}\sum_{n=0}^\infty\sum_{k=0}^\infty
\sum_{\cal O}N^{\cal O}_{n,k}\sum_{\vec{\nu}\in{\cal O}}
\mathrm{Li}_3(e^{2\pi i n\phi+2\pi i k\tau+i\vec{\nu}\cdot\vec{m}}).
\end{equation}
Here $\vec{\nu}$ runs over the weights on the Weyl orbit ${\cal O}$.
Partition functions $Z_n$ of $U(n)$ gauge theories
on ${1\over 2}K_3$ are then defined by
\be
{\cal F}(\phi,\tau,\vec{m})={\cal F}_{\mbox{\scriptsize classical}}
-\frac{1}{(2\pi i)^3}\sum_{n=1}^\infty
q^{n/2}Z_n(\vec{m};\tau)e^{2\pi i n\phi},\hskip3mm q=e^{2\pi i\tau}.
\label{prepob}\end{equation}
(An extra factor of $q^{n/2}$ has been introduced in the right-hand side 
so that $Z_n$ has a simpler modular property).
Prepotential (\ref{prepoa}),\,(\ref{prepob}) 
is also interpreted as the generating function for 
the number of $E$-string BPS states
or the holomorphic curves of ${1\over 2}K_3$.
Variable $u$ and $\phi$ are related to each other
by a mirror-type transformation of Seiberg--Witten theory.

It is known that the instanton amplitudes $Z_n(\vec{m};\tau)$
are Jacobi forms and 
possess good modular properties except for the ``anomaly'': 
i.e. they contain contributions of $E_2$, the Eisenstein series
with weight $2$ which possesses an anomalous transformation law.
Dependence of $Z_n$ on $E_2$ is determined
by the holomorphic anomaly equation \cite{MNW2}
\be
{\partial Z_n\over \partial E_2}={1\over 24}\sum_{m=1}^{n-1}m(n-m)Z_mZ_{n-m}.
\label{holanom}\end{equation}
Holomorphic anomaly represents the contributions of reducible connections
of gauge theory on ${1\over 2}K_3$ which possesses $b_2^+=1$.
 
It is known that the amplitude $Z_n$ also obeys the ``gap'' condition,
\be
N_{n,k}=0 \quad \mbox{for} \ \ k< n.
\label{gap}\end{equation}
Namely, holomorphic curves in the class $n[\Sigma]+k[E]$
exist only for $k\ge n$ in ${1\over 2}K_3$.
Amplitudes $Z_n$ have been determined for lower values of $n$  perturbatively 
in \cite{MNVW} by making use of
the holomorphic anomaly equation (\ref{holanom}) 
and gap conditions (\ref{gap}).

It is well-known that given a Seiberg--Witten curve
one can compute the amplitudes $Z_n$ and
generate the instanton expansion (\ref{prepoa})
by following the standard steps of the Seiberg--Witten theory.
In our previous paper we have taken an inverse procedure:
we first computed $Z_n$ up to sufficiently high-orders
in $n$ using holomorphic anomaly and 
gap condition and then used these data to 
determine the Seiberg--Witten curve ${\cal C}$. 
It turned out that functions $f(u;\tau,\vec{m})$ and
$g(u;\tau,\vec{m})$ are expressed in terms of
the characters of affine $E_8$ algebra up to level 4 and 6 \cite{ES}.
We have checked the consistency of our curve in various ways:
it possesses the correct modular properties and in particular 
reproduces the known five- and four-dimensional curves in the 
degenerate limit $R_6, R_5 \rightarrow 0$. 

In this paper we first study holomorphic sections of the
rational elliptic surface by making use of our curve ${\cal C}$. 
In the mathematical literature \cite{Shi}
it is known that meromorphic sections of an elliptic surface
form a lattice under a suitable
definition for their addition and their inner product.
In the case of a rational elliptic surface \cite{OS,FYY},
this lattice (Mordell--Weil lattice) 
coincides with the root lattice of $E_8$. 
In particular there exist 240 
holomorphic sections 
corresponding to non-zero roots of $E_8$. 

We first construct a holomorphic section explicitly which corresponds to a 
given root of $E_8$ by making use of affine $E_7$ characters. 
Then the rest of 240 sections can be simply obtained by the 
application of $E_8$ Weyl transformations to this section.

We shall show that the Wilson line parameters $\{m_i\}$ are identified as the
coordinates of the points on the elliptic fiber $E_{\infty}$ where holomorphic
sections intersect.
This is in fact the geometrical significance of the parameters $\{m_i\}$ 
suggested in the literature \cite{GMS,Moh}. 
We also present the calculation of sections of five-dimensional curves
for the sake of illustrations.

In the latter half of this paper we investigate
various specializations of our Seiberg--Witten curve.
First we study the relation between sets of special values of
Wilson line parameters
$\{m_i\}$ and the corresponding degenerations of elliptic fibration:
for instance, we find an unbroken symmetry $E_7\oplus A_1$
when all the parameters $m_i$ are set equal to
$\pi/2$ while we have an unbroken $E_8$ symmetry when they are all set to zero
$m_i=0$. In general
when one adjusts the parameters $\{m_i\}$ 
so that they preserve a symmetry under a subgroup of $E_8$,
elliptic fibration exhibits corresponding degenerations.
Degenerate fibers form $ADE$-type singularities in Kodaira's classification and
correspond precisely to the unbroken symmetry group.

Partial specification of Wilson line parameters in our curve
also produces interesting results.
As we mentioned already, our curve contains
a series of known Seiberg--Witten curves for
five- and four-dimensional gauge theories and
the latter are obtained by taking 
the degenerate limit of $T^2$
and sending some of the mass parameters to $\infty$.
In this way of reduction, however, we necessarily lose $SL(2,{\bf Z})$ symmetry
of $T^2$ and can not naturally 
recover the $SL(2,{\bf Z})$ symmetry of $N_f=4$ theory in four 
dimensions. 

In this paper we propose new type of reductions to four dimensions without
taking the degenerate limit of $T^2$ so that
the $SL(2,{\bf Z})$ symmetry is left intact.
In fact, for instance, 
by setting four of the parameters $\{m_i\}$ to
half-periods $(0,\pi,\pi+\pi\tau,\pi\tau)$ we obtain
the curve for the $SU(2)\ N_f=4$ gauge theory directly
from our six-dimensional curve.
As a remarkable by-product,
the $SO(8)$ triality of the $N_f=4$ theory is 
derived from the $SL(2,{\bf Z})$ symmetry
of the six-dimensional curve in a natural manner.

The organization of this paper is as follows:
In section 2 we recall the basic symmetry properties of our
six-dimensional curve (we call it as the $\aE_8$ curve)
and its five-dimensional counterparts ($E_n$ curves).
In section 3 we derive some holomorphic sections for
$E_n$ curves for the sake of illustrations.
In section 4 we present holomorphic sections for
the $\aE_8$ curve and identify the geometrical significance of
the Wilson line parameters.
In section 5 we present
examples which exhibit
the correspondence between special values of 
Wilson line parameters and patterns of degenerations of elliptic fibration.
In section 6 we discuss a new reduction to four dimensions and 
show how to obtain
the Seiberg--Witten curve
for the $SU(2)\ N_f=4$ gauge theory.
We also present a reduction to four-dimensional Donagi--Witten
theory \cite{DW}.
Section 7 is devoted to discussions.
There are three Appendices in this paper: in Appendix A
we present the explicit form of the $\aE_8$ curve and in Appendix B
we list those of the $E_n$ curves together with their holomorphic sections.
In Appendix C we also present a list of special configurations of Wilson line
parameters which lead to various unbroken subgroups of $E_8$.

\section{Basics of the $\hE_8$ Curve and $E_n$ Curves}

A rational elliptic surface possesses the structure of
an elliptic fibration over a base
${\bf P}^1$.
It can be described as a family of elliptic curves in the Weierstrass form 
\be \begin{array}{rcl}
y^2&=&4x^3-\left(a_0u^4+a_1u^3+a_2u^2+a_3u+a_4\right)x 
\\ & &-\left(b_0u^6+b_1u^5+b_2u^4+b_3u^3+b_4u^2+b_5u+b_6\right)\end{array}
\label{swcurve}\end{equation}
where $u$ denotes the coordinate for the base ${\bf P}^1$.
Curve (\ref{swcurve}) possesses $12$ moduli out of which
2 may be eliminated by the scaling and shift of the variable $u$
in agreement with 
$h^{1,1}=10$ for the rational elliptic surface. We choose a convention
$a_1=0$ in this paper. Actually
we can also fix the degree of freedom of the rescaling 
$(u,x,y)\to(Lu,L^2x,L^3y)$ so that we are left with 
nine parameters $\tau,\{m_i,i=1,2,\dots,8\}$ to parametrize the
coefficient functions $\{a_j,b_j\}$.
Their explicit form is given in Appendix A.

The functions $a_j(m_i;\tau),b_j(m_i;\tau)$ are expressed in terms of the
characters of the $\aE_8$ Weyl orbits at level $j$.
Thus the curve possesses manifest $\aE_8$
Weyl group symmetry and thereby we call it $\aE_8$ curve.
Strictly speaking
the curve exhibits three kinds of automorphisms:
the $E_8$ Weyl group symmetry,
the double periodicity in $\vec{m}$,
and the modular property in $\tau$.

First the curve is invariant under the following Weyl reflections
\be
\begin{array}{rcll}
\bullet\ m_i\Eqn{\leftrightarrow}m_j\quad &(i\not=j),\\
\bullet\ m_i\Eqn{\leftrightarrow}-m_j\quad &(i\not=j),\\
\bullet\ m_i\Eqn{\to} m_i-\frac{1}{4}\sum_{j=1}^8m_j\ &(i=1,\ldots,8).
\end{array}
\end{equation}
The entire $E_8$ Weyl group is generated by the combination of
these operations.
On the other hand the double periodicity in $\vec{m}$ takes the following form:
let $\vec{\alpha}$ be any vector of the root lattice $\Gamma_8$.
Then the $\aE_8$ curve is invariant under
\ba
\bullet\ \vec{m}\Eqn{\to}\vec{m}+2\pi\vec{\alpha},\\
\bullet\ \vec{m}\Eqn{\to}\vec{m}+2\pi\tau\vec{\alpha}\quad\mbox{with}
\quad (u,x,y)\to (Lu,L^2x,L^3y)\no \\ \Eqn{}\hspace{10em}\quad\mbox{where}
\quad L=e^{-i\vec{m}\cdot\vec{\alpha}}q^{-\frac{1}{2}|\vec{\alpha}|^2}.
\ea
Curve also remains invariant under the action of
the $SL(2,{\bf Z})$ symmetry generated by
\ba
\bullet\ \tau\Eqn{\to}\tau+1,\\
\bullet\ \tau\Eqn{\to}-\frac{1}{\tau},
  \ \vec{m}\to\frac{\vec{m}}{\tau}\quad\mbox{with}\quad
(u,x,y)\to(\tau^{-6}Lu,\tau^{-10}L^2x,\tau^{-15}L^3y)\no\\
\Eqn{}\hspace{15em}\mbox{where}\quad L=e^{\frac{i}{4\pi\tau}|\vec{m}|^2}.
\ea

Next let us review the reduction of our $\aE_8$ curve
down to the five-dimensional situation.
By taking the limit $\mathrm{Im}\tau\to\infty\ (q\to 0)$
we obtain a curve whose coefficients are written in terms of
characters of the finite-dimensional $E_8$ algebra.
We call it the $E_8$ curve.
There exists another `$E_8$ curve'
in four-dimensional theory \cite{MN2} which can be obtained
by taking a further degeneration \cite{ES}.
Properties of the latter $E_8$ curve
have been studied in \cite{MN2,NTY},
so we concentrate on the former.

Let us introduce some notations associated with the $E_8$ algebra.
Let $\vec{\Lambda}$ be some dominant weight
and $\vec{\mu}_1,\ldots,\vec{\mu}_8$
be the fundamental weights of $E_8$.
We then introduce a notation
\be
\vec{\Lambda}=\Biggl[
\begin{array}{ccccccc}
 & & n_2 & & & &\\
n_1 &n_3 &n_4 &n_5 &n_6 &n_7 &n_8
\end{array}\Biggr]=\sum_{i=1}^8 n_i\vec{\mu}_i,
\quad n_i\in {\bf Z}_{\ge 0}
\end{equation}
where $\{n_i\}$ denote Dynkin indices of $\vec{\Lambda}$.
We fix a labeling of the fundamental weights by
placing eight indices at eight nodes of the Dynkin diagram.
Next we define a character
for an irreducible representation $\mathcal{R}$
of highest weight $\vec{\Lambda}$ by
\be
\chi_\mathcal{R}^{E_8}(m_i)=
\chEei{n_1}{n_2}{n_3}{n_4}{n_5}{n_6}{n_7}{n_8}(m_i)
\equiv\sum_{\vec{\nu}\in\mathcal{R}}e^{i\vec{m}\cdot\vec{\nu}}
\end{equation}
where $\vec{\nu}$ runs over all weights of
representation $\mathcal{R}$. 
The variable $\vec{m}$ takes its values on the $\bf C$-extended
root space. Then characters of fundamental representations are written as
\ba
&&
\chi_1^{E_8}=\chEei{1}{0}{0}{0}{0}{0}{0}{0}, \ \
\chi_2^{E_8}=\chEei{0}{1}{0}{0}{0}{0}{0}{0}, \ \
\chi_3^{E_8}=\chEei{0}{0}{1}{0}{0}{0}{0}{0}, \ \
\no\\ &&
\chi_4^{E_8}=\chEei{0}{0}{0}{1}{0}{0}{0}{0},\ \ 
\chi_5^{E_8}=\chEei{0}{0}{0}{0}{1}{0}{0}{0}, \ \
\chi_6^{E_8}=\chEei{0}{0}{0}{0}{0}{1}{0}{0}, \ \
\no\\ &&
\chi_7^{E_8}=\chEei{0}{0}{0}{0}{0}{0}{1}{0}, \ \
\chi_8^{E_8}=\chEei{0}{0}{0}{0}{0}{0}{0}{1}. \hspace{3em}
\ea
With this preparation we can now express the $E_8$ curve
in terms of these characters.
The explicit form is given in Appendix B.
We have applied a shift in $x$ for the compactness of the 
expression.
Weierstrass form can be immediately recovered by shifting $x$
to cancel the $x^2$ term.

The $E_7$ curve is obtained from the $E_8$ curve by
decomposing the $E_8$ characters into those of
$E_7\times U(1)$ and factoring out the $U(1)$ parts.
The operation is carried out by the following limit
\ba
\Eqn{}
(\chi_1^{E_8},\chi_2^{E_8},\chi_3^{E_8},\chi_4^{E_8},
 \chi_5^{E_8},\chi_6^{E_8},\chi_7^{E_8},\chi_8^{E_8})\no\\
\Eqn{}\hspace{1em}\to
(L^2\chi_1^{E_7},L^3\chi_2^{E_7},L^4\chi_3^{E_7},L^6\chi_4^{E_7},
 L^5\chi_5^{E_7},L^4\chi_6^{E_7},L^3\chi_7^{E_7},L^2),\\
\Eqn{}(u,x,y)\to(Lu,L^2x,L^3y)\quad\mbox{with}\quad L\to\infty
\ea
where $\chi_1^{E_7},\ldots,\chi_7^{E_7}$ denote
characters of fundamental representations of $E_7$
\ba
\Eqn{}
\chi_1^{E_7}=\chEse{1}{0}{0}{0}{0}{0}{0}, \ \
\chi_2^{E_7}=\chEse{0}{1}{0}{0}{0}{0}{0}, \ \
\chi_3^{E_7}=\chEse{0}{0}{1}{0}{0}{0}{0}, \ \
\chi_4^{E_7}=\chEse{0}{0}{0}{1}{0}{0}{0},\no\\
\Eqn{}
\chi_5^{E_7}=\chEse{0}{0}{0}{0}{1}{0}{0}, \ \
\chi_6^{E_7}=\chEse{0}{0}{0}{0}{0}{1}{0}, \ \
\chi_7^{E_7}=\chEse{0}{0}{0}{0}{0}{0}{1}.
\ea

The $E_6$ curve is obtained from the $E_7$ curve in a similar manner as
\ba
&&
(\chi_1^{E_7},\chi_2^{E_7},\chi_3^{E_7},\chi_4^{E_7},
 \chi_5^{E_7},\chi_6^{E_7},\chi_7^{E_7})\no\\
&&\hspace{2em}\to
(L^2\chi_1^{E_6},L^3\chi_2^{E_6},L^4\chi_3^{E_6},L^6\chi_4^{E_6},
 L^5\chi_5^{E_6},L^4\chi_6^{E_6},L^3),\\
&&(u,x,y)\to(Lu,L^2x,L^3y)\quad\mbox{with}\quad L\to\infty
\ea
where $\chi_1^{E_6},\ldots,\chi_6^{E_6}$ denote
characters of fundamental representations of $E_6$
\ba
&&
\chi_1^{E_6}=\chEsi{1}{0}{0}{0}{0}{0}, \ \
\chi_2^{E_6}=\chEsi{0}{1}{0}{0}{0}{0}, \ \
\chi_3^{E_6}=\chEsi{0}{0}{1}{0}{0}{0}, \ \
\chi_4^{E_6}=\chEsi{0}{0}{0}{1}{0}{0},\no\\
&&
\chi_5^{E_6}=\chEsi{0}{0}{0}{0}{1}{0}, \ \
\chi_6^{E_6}=\chEsi{0}{0}{0}{0}{0}{1}.
\ea
Lower $E_n$ curves are obtained in a similar fashion.

Five-dimensional $E_n$ curves exhibit $E_n$ Weyl group symmetry
and a single periodicity
on the mass parameters $\{m_i\}$. In fact the curves remain invariant under
\be\label{E_n-periodicty}
\vec{m}\to\vec{m}+2\pi\vec{\alpha}
\end{equation}
where $\vec{\alpha}$ is any vector of $E_n$ root lattice.

\section{Holomorphic Sections for $E_n$ Curves}

Let us begin with the case of $E_6$ curve in order to illustrate our method.
The $E_6$ curve has the form
\begin{eqnarray}\label{E6curve}
y^2 \Eqn{=} 4x^3
  +\Bigl(-u^2+4\chi_1^{E_6}\Bigr)x^2
  +\Bigl((2\chi_2^{E_6}-12)u+(4\chi_3^{E_6}-4\chi_6^{E_6})\Bigr)x\no\\
\Eqn{}+4u^3+4\chi_6^{E_6}u^2+(4\chi_5^{E_6}-4\chi_1^{E_6})u
  +(4\chi_4^{E_6}-\chi_2^{E_6}\chi_2^{E_6}).
\end{eqnarray}
A holomorphic section is given by a pair of polynomials 
$(x=x(u),y=y(u))$ in $u$ since $x,y$ have to be holomorphic
on the entire $u$-plane.
To cancel the $u^2$ term in the coefficient of $x^2$ in (\ref{E6curve}),
$x(u)$ must be at least linear in $u$.
So we assume the following form for a section
\be\label{assum-sec}
\sigma_{27}\equiv\left\{
\begin{array}{l}
x(u)=x_1u+x_2,\\
y(u)=y_1u^2+y_2u+y_3.
\end{array}
\right.
\end{equation}
By substituting this ansatz into the $E_6$ curve (\ref{E6curve}),
we obtain a polynomial of degree 4 in $u$ vanishing identically.
Coefficients of $u^k\ k=0,\ldots,4$ provide five equations
for five unknowns $x_1,x_2,y_1,y_2,y_3$.
By eliminating $y_1,y_2,y_3,x_2$ we obtain
two sets of 27th-degree equations in $x_1$.
Solutions of these equations give
holomorphic sections corresponding to
${\bf 27}$ and ${\bf \overline{27}}$ representations of $E_6$.

To obtain the explicit expression of sections,
we first note that every weight of ${\bf 27}$
preserves the $E_5(=D_5)$ Weyl group symmetry.
So let us consider the $E_6\to E_5\times U(1)$ decomposition.
Fundamental characters
branch as:
\begin{eqnarray}
\label{E6E5-1}
\chi_1^{E_6}\Eqn{=}\chi_1^{E_5} \Lambda^2
  +\chi_2^{E_5} \Lambda^{-1}+\Lambda^{-4},\\
\chi_2^{E_6}\Eqn{=}\chi_2^{E_5} \Lambda^3
  +(\chi_3^{E_5}+1)+\chi_5^{E_5} \Lambda^{-3},\\
\chi_3^{E_6}\Eqn{=}\chi_3^{E_5} \Lambda^4+\chi_1^{E_5} \chi_2^{E_5} \Lambda
  +(\chi_1^{E_5}+\chi_4^{E_5}) \Lambda^{-2}+\chi_2^{E_5} \Lambda^{-5},\\
\chi_4^{E_6}\Eqn{=}\chi_4^{E_5} \Lambda^6+\chi_2^{E_5} \chi_3^{E_5} \Lambda^3
  +(\chi_1^{E_5} \chi_4^{E_5}+\chi_3^{E_5})
  +\chi_3^{E_5} \chi_5^{E_5} \Lambda^{-3}+\chi_4^{E_5}
  \Lambda^{-6},\no\\ \\
\chi_5^{E_6}\Eqn{=}\chi_5^{E_5}\Lambda^5+(\chi_1^{E_5}+\chi_4^{E_5})\Lambda^2
  +\chi_1^{E_5} \chi_5^{E_5} \Lambda^{-1}+\chi_3^{E_5} \Lambda^{-4},\\
\label{E6E5-6}
\chi_6^{E_6}\Eqn{=} \Lambda^4+\chi_5^{E_5} \Lambda+\chi_1^{E_5} \Lambda^{-2}
\end{eqnarray}
where $\chi_1^{E_5},\ldots,\chi_5^{E_5}$ denote
the $E_5$ fundamental characters
\ba&&
\chi_1^{E_5}=\chEfi{1}{0}{0}{0}{0}, \ \
\chi_2^{E_5}=\chEfi{0}{1}{0}{0}{0}, \ \
\chi_3^{E_5}=\chEfi{0}{0}{1}{0}{0}, \no\\ &&
\chi_4^{E_5}=\chEfi{0}{0}{0}{1}{0}, \ \
\chi_5^{E_5}=\chEfi{0}{0}{0}{0}{1}.
\ea
$\Lambda\equiv e^{i\lambda}$ represents the $U(1)$ character with unit charge
where $\lambda$ is the projection of $\vec{m}$
in the direction orthogonal to $E_5$.
(We have suitably normalized $\lambda$ so that all the weights
have integer $U(1)$ charges.)
(\ref{E6E5-1})--(\ref{E6E5-6})
show the standard branching of $E_6$ representations;
for example, (\ref{E6E5-1}) represents the decomposition
${\bf 27}={\bf 10}_2+{\bf 16}_{-1}+{\bf 1}_{-4}$.
With these decompositions
it is not difficult to find the following solution
for the simultaneous algebraic equations
\begin{eqnarray}
x(u)\Eqn{=}-\Lambda^4 u-(\Lambda^8
  +\chi_1^{E_5}\Lambda^2+\Lambda^{-4}),\\
\label{E627y}
y(u)\Eqn{=}i\Lambda^4 u^2+i(3 \Lambda^8
  +\chi_1^{E_5} \Lambda^2-\Lambda^{-4})u\no\\
\Eqn{}+i(2 \Lambda^{12}+2 \chi_1^{E_5} \Lambda^6
  -\chi_2^{E_5} \Lambda^3
  +\chi_3^{E_5}+1-\chi_5^{E_5} \Lambda^{-3}).
\end{eqnarray}
Explicit dependence of $\Lambda$ and $\chi_j^{E_5}$
on the parameters $\{m_i\}$ is fixed in 27 ways
corresponding to each weight of the representation ${\bf 27}$. 
Thus the above formula compactly represents 27 holomorphic sections
for $E_6$ curve.

Also we can immediately obtain the sections corresponding to 
${\bf \overline{27}}$ by reflecting the overall sign of
the solution $y(u)$, i.e. right-hand side of (\ref{E627y}).

Next let us consider sections of the form
\ba
\sigma_{72}\equiv\left\{
\begin{array}{l}
x(u)=x_0u^2+x_1u+x_2,\\
y(u)=y_0u^3+y_1u^2+y_2u+y_3.
\end{array}
\right.
\ea
By eliminating variables $y_0,y_1,y_2,y_3,x_2,x_1$
one obtains a 36th-degree equation in $x_0$
which represents 36 pairs of
roots of the adjoint representation of $E_6$.
In this case, $A_5$ is the surviving symmetry
orthogonal to a root of $E_6$,
so we consider the following decomposition:
\begin{eqnarray}
\chi_1^{E_6}\Eqn{=}2 \chi_1^{A_5} \cos\lambda+\chi_4^{A_5},\\
\chi_2^{E_6}\Eqn{=}2 \cos 2\lambda+2 \chi_3^{A_5} \cos\lambda
  +\chi_1^{A_5} \chi_5^{A_5},\\
\chi_3^{E_6}\Eqn{=}2 \chi_2^{A_5} \cos 2\lambda
  +2 \chi_1^{A_5} \chi_4^{A_5} \cos\lambda
  +(\chi_1^{A_5}\chi_1^{A_5}+\chi_3^{A_5} \chi_5^{A_5}-\chi_2^{A_5}),\\
\chi_4^{E_6}\Eqn{=}2 \chi_3^{A_5} \cos 3\lambda
  +2 \chi_2^{A_5} \chi_4^{A_5} \cos 2\lambda
  +2 \chi_1^{A_5} \chi_3^{A_5} \chi_5^{A_5} \cos\lambda\no\\
\Eqn{}+(\chi_1^{A_5}\chi_1^{A_5} \chi_4^{A_5}
  +\chi_2^{A_5} \chi_5^{A_5}\chi_5^{A_5}
  -\chi_1^{A_5} \chi_5^{A_5}+\chi_3^{A_5}\chi_3^{A_5}
  -2 \chi_2^{A_5} \chi_4^{A_5}+1),\no\\ \\
\chi_5^{E_6}\Eqn{=}2 \chi_4^{A_5} \cos 2\lambda
  +2 \chi_2^{A_5} \chi_5^{A_5} \cos\lambda
  +(\chi_1^{A_5} \chi_3^{A_5}+\chi_5^{A_5}\chi_5^{A_5}-\chi_4^{A_5}),\\
\chi_6^{E_6}\Eqn{=}2 \chi_5^{A_5} \cos\lambda+\chi_2^{A_5}
\end{eqnarray}
where $\chi_1^{A_5},\ldots,\chi_5^{A_5}$ denote
$A_5$ fundamental characters
\ba &&
\chi_1^{A_5}=\chAfi{1}{0}{0}{0}{0}, \ \
\chi_2^{A_5}=\chAfi{0}{1}{0}{0}{0}, \ \
\chi_3^{A_5}=\chAfi{0}{0}{1}{0}{0}, \no \\ &&
\chi_4^{A_5}=\chAfi{0}{0}{0}{1}{0}, \ \
\chi_5^{A_5}=\chAfi{0}{0}{0}{0}{1}.
\ea
$\lambda$ is the projection of $\vec{m}$
orthogonal to $A_5$.
Then we find the following solution:
\begin{eqnarray}
x(u)\Eqn{=}\frac{1}{4\sin^2\lambda}
  \left[u^2+(2 \chi_5^{A_5} \cos\lambda)\, u
  -4\chi_4^{A_5}\sin^2\lambda+\chi_5^{A_5}\chi_5^{A_5}\right],\\
y(u)\Eqn{=}\frac{1}{4\sin^3\lambda}\times\no\\
\Eqn{}\left[(\cos\lambda) \,u^3
  +\Bigl(\chi_5^{A_5} \cos 2\lambda+2 \chi_5^{A_5}\Bigr) u^2\right.\no\\
\Eqn{}\ 
  +\Bigl(\chi_4^{A_5} \cos 3\lambda-2 \chi_1^{A_5} \cos 2\lambda
  +(3 \chi_5^{A_5}\chi_5^{A_5}-\chi_4^{A_5})
  \cos\lambda+2 \chi_1^{A_5}\Bigr) u\no\\
\Eqn{}\ 
  -\cos 5\lambda+\chi_3^{A_5} \cos 4\lambda
  +(-\chi_1^{A_5} \chi_5^{A_5}+3) \cos 3\lambda
\no\\ \Eqn{} \  
+(2 \chi_4^{A_5} \chi_5^{A_5}-4 \chi_3^{A_5}) \cos 2\lambda 
  +(\chi_1^{A_5} \chi_5^{A_5}-2) \cos\lambda
\no\\ \Eqn{} \ 
 +(\chi_5^{A_5}\chi_5^{A_5}\chi_5^{A_5}
    -2\chi_4^{A_5} \chi_5^{A_5}+3 \chi_3^{A_5})\Bigr].
\end{eqnarray}
This expression represents $72$ sections corresponding to
the roots of $E_6$.

In Appendix B we present holomorphic sections
for the $E_7$ and $E_8$ curve.

\section{Holomorphic Sections for the $\hE_8$ Curve}

In the case of $\aE_8$ curve the derivation of holomorphic sections become
much more complex than $E_n$ curves.
Thus instead of solving algebraic equations 
in the generic case of 8 parameters we first consider the 
simpler case of only 2 non-vanishing Wilson line parameters. We then recover 
the general case by making use of the modular invariance and $\aE_7$ symmetry.

Let us consider sections of the form
\be\label{6dsection}
\sigma_{240}\equiv\left\{
\begin{array}{l}
x(u)=x_0 u^2+x_1 u+x_2,\\
y(u)=y_0 u^3+y_1 u^2+y_2 u+y_3.
\end{array}
\right.
\end{equation}
In order to maintain the modular invariance,
coefficients $x_k,y_k$ must transform as
\ba
\label{Ttransf-x}
x_k(\vec{m};\tau+1)\Eqn{=}x_k(\vec{m};\tau),\\
y_k(\vec{m};\tau+1)\Eqn{=}y_k(\vec{m};\tau)
\ea
and
\ba
x_k\left(\frac{\vec{m}}{\tau};-\frac{1}{\tau}\right)
\Eqn{=}\tau^{2-6k}e^{\frac{ik}{4\pi\tau}|\vec{m}|^2}
  x_k(\vec{m};\tau),\\
\label{Stransf-y}
y_k\left(\frac{\vec{m}}{\tau};-\frac{1}{\tau}\right)
\Eqn{=}\tau^{3-6k}e^{\frac{ik}{4\pi\tau}|\vec{m}|^2}
  y_k(\vec{m};\tau).
\ea

We assume that each section preserves
an $E_7$ Weyl symmetry and can be expressed
in terms of $\aE_7$ characters.
In the following $\aE_7$ characters are presented as
explicit functions of $m_i$ and $\tau$.
We introduce an orthogonal coordinate basis $\{\vec{\bf e}_j\}\ j=1,\ldots,8$
for the $E_8$ root lattice so that
\ba
\left\{
\begin{array}{cc}
\pm\vec{\bf e}_i\pm\vec{\bf e}_j & i\ne j,\\
\frac{1}{2}(\pm\vec{\bf e}_1\pm\vec{\bf e}_2\pm\cdots\pm\vec{\bf e}_8)
  &\mbox{with even number of $+$'s}
\end{array}
\right.
\ea
form the 240 roots as usual.
Corresponding to each one of these roots there exists a
holomorphic section which
is invariant under $E_7$ Weyl group orthogonal to the root.
As an illustration we consider a root
\be
\vec{\bf e}_7+\vec{\bf e}_8
\end{equation}
and the $E_7$ root lattice orthogonal to it.
$\aE_7$ has two Weyl orbits at level one
whose characters are expressed as
\ba
w^{\aE_7}_{\rm b}(m_i;\tau)\Eqn{=}
\frac{1}{2}\left[
\left(\prod_{j=1}^6\varth_1(m_j|\tau)+\prod_{j=1}^6\varth_2(m_j|\tau)\right)
\varth_2(2m_-|2\tau)\right.\no\\
&&\hspace{1.35em} \left.
+\left(\prod_{j=1}^6\varth_3(m_j|\tau)+\prod_{j=1}^6\varth_4(m_j|\tau)\right)
\varth_3(2m_-|2\tau)
\right],\\
w^{\aE_7}_{\rm f}(m_i;\tau)\Eqn{=}
\frac{1}{2}\left[
\left(-\prod_{j=1}^6\varth_1(m_j|\tau)+\prod_{j=1}^6\varth_2(m_j|\tau)\right)
\varth_3(2m_-|2\tau)\right.\no\\
&&\hspace{1.35em} \left.
+\left(\prod_{j=1}^6\varth_3(m_j|\tau)-\prod_{j=1}^6\varth_4(m_j|\tau)\right)
\varth_2(2m_-|2\tau)
\right]
\ea
where $m_-=(m_7-m_8)/2$.
Note that the level-one $\aE_8$ Weyl orbit character
\be
w^{\aE_8}_{\rm b}(m_i;\tau)=
P(m_i;\tau)=\frac{1}{2}\sum_{\ell=1}^4\prod_{j=1}^8\varth_\ell(m_j|\tau)
\end{equation}
decomposes as
\be
w^{\aE_8}_{\rm b}(m_i;\tau)=
 w^{\aA_1}_{\rm b}(m_+;\tau)w^{\aE_7}_{\rm b}(m_i;\tau)
+w^{\aA_1}_{\rm f}(m_+;\tau)w^{\aE_7}_{\rm f}(m_i;\tau)
\end{equation}
where the branching functions are the level-one characters of $\aA_1$
\be
w^{\aA_1}_{\rm b}(m_+;\tau)=\varth_3(2m_+|2\tau),\qquad
w^{\aA_1}_{\rm f}(m_+;\tau)=\varth_2(2m_+|2\tau)
\end{equation}
and $m_+=(m_7+m_8)/2$.

With the help of the transformation rules
(\ref{Ttransf-x})--(\ref{Stransf-y}) and the $\aE_7$ symmetry,
we need not solve the simultaneous algebraic equations
with the most generic Wilson line parameters.
It turns out that the case of
two non-zero parameters is enough to 
determine general sections completely.
On the other hand, it is known that sections with two non-zero parameters
can be relatively easily obtained \cite{MNW1}.
By rewriting these sections in terms of an $\aE_8$ invariant coordinate 
and recovering all the Wilson line parameters,
we obtain holomorphic sections for the $\aE_8$ curve.
Holomorphic section corresponding to
the root $\vec{\bf e}_7+\vec{\bf e}_8$ is given explicitly as
\ba\label{6dx0}
x_0=\wp(2m_+),
\ea
\be
x_1=
 C_{\rm b}(m_+,\tau)w^{\aA_1}_{\rm b}(m_+;\tau)w^{\aE_7}_{\rm b}(m_i;\tau)
+C_{\rm f}(m_+,\tau)w^{\aA_1}_{\rm f}(m_+;\tau)w^{\aE_7}_{\rm f}(m_i;\tau)
\end{equation}
where
\ba
C_{\rm b}(m_+,\tau)\Eqn{=}
-\frac{{\varth_2}^8({\varth_3}^4+{\varth_4}^4)}{4\eta^{24}E_4}\wp(2m_+)
  +\frac{{\varth_2}^8}{48\eta^{24}}
  +\frac{\varth_2(2m_+|2\tau)^4}{\eta^{18}\varth_1(2m_+|\tau)^2},\no\\
  \\
C_{\rm f}(m_+,\tau)\Eqn{=}
-\frac{{\varth_2}^8({\varth_3}^4+{\varth_4}^4)}{4\eta^{24}E_4}\wp(2m_+)
  +\frac{{\varth_2}^8}{48\eta^{24}}
  +\frac{\varth_3(2m_+|2\tau)^4}{\eta^{18}\varth_1(2m_+|\tau)^2},\no\\
\ea
\ba
x_2\Eqn{=}{1\over 2}D(m_+,\tau)
\left(w^{\aA_1}_{\rm b}(2m_+;2\tau)
  +w^{\aA_1}_{\rm f}(2m_+;2\tau)\right)\no\\\Eqn{}
\left(w^{\aE_7}_{\rm b}(2m_i;2\tau)
  +w^{\aE_7}_{\rm f}(2m_i;2\tau)\right)\no\\
\Eqn{+}{1\over 2}D(m_+,\tau+1)
\left(w^{\aA_1}_{\rm b}(2m_+;2\tau)
  -w^{\aA_1}_{\rm f}(2m_+;2\tau)\right)\no\\\Eqn{}
\left(w^{\aE_7}_{\rm b}(2m_i;2\tau)
  -w^{\aE_7}_{\rm f}(2m_i;2\tau)\right)\no\\
\Eqn{+}\frac{\tau^{14}}{2^4}D\Bigl(\frac{m_+}{\tau},-\frac{1}{\tau}\Bigr)
w^{\aA_1}_{\rm b}\Bigl(m_+;\frac{\tau}{2}\Bigr)
w^{\aE_7}_{\rm b}\Bigl(m_i;\frac{\tau}{2}\Bigr)\no\\
\Eqn{+}\frac{\tau^{14}}{2^4}D\Bigl(\frac{m_+}{\tau},-\frac{1}{\tau}+1\Bigr)
w^{\aA_1}_{\rm f}\Bigl(m_+;\frac{\tau}{2}\Bigr)
w^{\aE_7}_{\rm f}\Bigl(m_i;\frac{\tau}{2}\Bigr)\no\\
\Eqn{+}\frac{(\tau+1)^{14}}{2^4}
  D\Bigl(\frac{m_+}{\tau+1},-\frac{1}{\tau+1}\Bigr)
w^{\aA_1}_{\rm b}\Bigl(m_+;\frac{\tau+1}{2}\Bigr)
w^{\aE_7}_{\rm b}\Bigl(m_i;\frac{\tau+1}{2}\Bigr)\no\\
\Eqn{+}\frac{(\tau+1)^{14}}{2^4}
  D\Bigl(\frac{m_+}{\tau+1},-\frac{1}{\tau+1}+1\Bigr)
w^{\aA_1}_{\rm f}\Bigl(m_+;\frac{\tau+1}{2}\Bigr)
w^{\aE_7}_{\rm f}\Bigl(m_i;\frac{\tau+1}{2}\Bigr)\no\\
\ea
where
\ba
D(m_+,\tau)\Eqn{=}
\frac{{\varth_3}^4{\varth_4}^4}{{\eta}^{36}{E_4}^2}
\left[({\varth_3}^{12}-28{\eta}^{12})\wp (2m_+)
  +\frac{1}{12}E_4(2{\varth_2}^8+{\varth_3}^8)\right].
\ea
Coefficient functions $y_k$ can be expressed in terms of
the above $\{x_k\}$ and coefficients $\{a_k,b_k\}$ of the $\aE_8$ curve as
\ba\label{6dy0}
y_0\Eqn{=}\wp'(2m_+),\\[1ex]
y_1\Eqn{=}\frac{1}{y_0}\left(6x_1{x_0}^2-\frac{1}{2}a_0x_1
  -\frac{1}{2}b_1\right)
,\\[1ex]
y_2\Eqn{=}\frac{1}{y_0}\left(
  -\frac{1}{2}{y_1}^2 +6 x_2 {x_0}^2+ 6 {x_1}^2x_0
  -\frac{1}{2}a_0x_2
  -\frac{1}{2}a_2 x_0-\frac{1}{2} b_2\right),\\[1ex]
\label{6dy3}
y_3\Eqn{=}\frac{1}{y_0}\left(
  -y_2 y_1+12 x_2 x_1 x_0 + 2 {x_1}^3
  - \frac{1}{2} a_2 x_1-\frac{1}{2}a_3 x_0 -\frac{1}{2}b_3\right).
\ea
Here $\wp(z)$ is the Weierstrass $\wp$-function
\be 
\wp(z)
=\frac{1}{z^2}+\!\!
\sum_{m,n\in{\bf Z}^2_{\ne (0,0)}}\!\!\left[\frac{1}{(z-\Omega_{m,n})^2}
  -\frac{1}{{\Omega_{m,n}}^2}\right],
\quad \Omega_{m,n}=2\pi(m+n\tau)
\end{equation}
and $\wp'(z)=\frac{\partial}{\partial z}\wp(z)$.
We have verified the above formulas by substituting them into the $\aE_8$ curve
and checking the equation
order by order in $q$ for a sufficiently high degree.

All the other holomorphic sections
corresponding to 240 roots of $E_8$
are immediately obtained
by the application of $E_8$ Weyl transformations.
Moreover, all the meromorphic sections
may also be generated from the holomorphic ones
if one uses the addition law of Mordell--Weil lattice.
The lattice is constructed by the following addition rule
and identification of the zero element:
Given two sections $S^{(i)}=(x^{(i)}(u),y^{(i)}(u))\ i=1,2$ 
of an elliptic fibration $y^2=4x^3-f(u)x-g(u)$
we can construct a 3rd section $S^{(3)}=(x^{(3)}(u),y^{(3)}(u))$ by
\ba
x^{(3)}\Eqn{=}-x^{(1)}-x^{(2)}
  +\frac{1}{4}\left(\frac{y^{(2)}-y^{(1)}}{x^{(2)}-x^{(1)}}\right)^2,\\
y^{(3)}\Eqn{=}
\frac{(x^{(3)}-x^{(2)})y^{(1)}-(x^{(3)}-x^{(1)})y^{(2)}}{x^{(2)}-x^{(1)}}.
\ea
This defines an addition law
$S^{(3)}=S^{(1)}+S^{(2)}$ of the Mordell--Weil lattice.
The zero element of the lattice is defined
by the zero section $(x(u)=\infty,y(u)=\infty)$ and 
the inverse of a section $(x(u),y(u))$ is given by $(x(u),-y(u))$.

For example, let us denote the section (\ref{6dsection})
with coefficients (\ref{6dx0})--(\ref{6dy3})
as $S_{\vec{\bf e}_7+\vec{\bf e}_8}$
and construct sections
$S_{\vec{\bf e}_6-\vec{\bf e}_7},\,S_{\vec{\bf e}_6+\vec{\bf e}_8}$
by Weyl reflections.
Then one can verify that
$S_{\vec{\bf e}_7+\vec{\bf e}_8}$
and $S_{\vec{\bf e}_6-\vec{\bf e}_7}$
actually add up to
$S_{\vec{\bf e}_6+\vec{\bf e}_8}$
by the above addition law.
Meromorphic sections correspond to weights with length-squared greater than 2
and can be generated by repeated use of the addition law.

The explicit form of holomorphic sections
tells us the geometrical meaning of the Wilson line parameters.
We first note that if we divide the $\aE_8$ curve by $u^6$
and take the limit $u \rightarrow \infty$, it is reduced to
\be
\left({y\over u^3}\right)^2=4\left({x\over u^2}\right)^3-a_0\,
\left({x\over u^2}\right)-b_0
=4\left({x\over u^2}\right)^3
-{E_4(\tau)\over 12}\left({x\over u^2}\right)-{E_6(\tau)\over 216}.
\end{equation}
This describes the fiber $E_{\infty}$ at $u=\infty$ with modulus $\tau$.
Then the leading coefficients $(x_0,y_0)$ of
the section $(x(u)=x_0u^2+\cdots,y(u)=y_0u^3+\cdots)$ 
are identified as the coordinates on $E_{\infty}$
\be
y_0^2=4x_0^3-{E_4(\tau)\over 12}x_0-{E_6(\tau)\over 216}.
\end{equation}
As we see in (\ref{6dx0}),\,(\ref{6dy0}), $(x_0,y_0)$ are
in fact expressed by the
Weierstrass function and its derivative.

We note that the argument of $\wp$-functions in (\ref{6dx0}),\,(\ref{6dy0})
can be written as
\be
2m_+=m_7+m_8=(\vec{\bf e}_7+\vec{\bf e}_8)\cdot\vec{m}.
\end{equation}
In general we find that a section corresponding to
a root $\vec{\alpha}$ intersects the fiber $E_{\infty}$
at $z=\vec{\alpha}\cdot\vec{m}$
\be\label{uinfsection}
(x_0,y_0)=(\wp(\vec{\alpha}\cdot\vec{m}),\wp'(\vec{\alpha}\cdot\vec{m})).
\end{equation}
Therefore the Wilson line parameters are identified as the
coordinates on the fiber at $u=\infty$. See Figure \ref{EF}.
This is in fact the identification suggested in the literature \cite{GMS,Moh}.
In our previous work \cite{ES} the precise dependence of
our Seiberg--Witten curve on the parameters
$\{m_i\}$ has been determined so that they reproduce
the correct mass dependence of the
lower-dimensional curves after reduction. We have now confirmed that they also 
possess the correct geometrical significance in six dimensions.

\section{Adjusting Parameters to Special Values}

Let us now investigate
various specializations of our Seiberg--Witten curve.
We study the relation between sets of special values of Wilson line parameters
$\{m_i\}$ and the corresponding degenerations of elliptic fibration.
In general when one adjusts the parameters $\{m_i\}$ 
so that they preserve a symmetry under a subgroup of $E_8$,
elliptic fibration exhibits corresponding degeneration.
Degenerate fibers form $ADE$-type singularities in Kodaira's classification and
correspond precisely to the unbroken symmetry group.
In this section we will demonstrate this correspondence in various cases.

As is well-known, we can identify the types of singularities
by the behaviors of the
functions $f,g$ and the discriminant
\be
y^2=4x^3-f(u)x-g(u), \hskip4mm \Delta=f^3-27g^2
\end{equation}
near the singularity as shown in Table \ref{SingFibers}.
\begin{table}[h]
\caption{Singular elliptic fibers\label{SingFibers}}
\[
\renewcommand{\arraystretch}{0.8}
\begin{array}{|c|c|c|c|c|}\hline
\mbox{order($f$)}&\mbox{order($g$)}&\mbox{order($\Delta$)}
  &\mbox{fiber-type}&\mbox{singularity-type}\\ \hline\hline
\ge 0&\ge 0&0&\mbox{smooth}&\mbox{---}\\ \hline
0&0&n&I_n&A_{n-1}\\ \hline
\ge 1&1&2&II&\mbox{---}\\ \hline
1&\ge 2&3&III&A_1\\ \hline
\ge 2&2&4&IV&A_2\\ \hline
2&\ge 3&n+6&I_n^\ast&D_{n+4}\\ \hline
\ge 2&3&n+6&I_n^\ast&D_{n+4}\\ \hline
\ge 3&4&8&IV^\ast&E_6\\ \hline
3&\ge 5&9&III^\ast&E_7\\ \hline
\ge 4&5&10&II^\ast&E_8\\ \hline
\end{array}
\]
\end{table}

In subsequent discussions the
patterns of degenerations of $\aE_8$ curve are
totally independent of the value of $\tau$.
So we fix $\tau$ to $i\infty$ and consider the five-dimensional $E_8$ curve.

\noindent
$\bullet\ E_8$\\
When we set all $\{m_i\}$ to $0$,
the whole $E_8$ Weyl group symmetry is preserved.
In this limit, each character becomes 
equal to the dimension of the corresponding representation
\be
\begin{array}{llll}
\chi_1=3875,\ &
\chi_2=147250,\ &
\chi_3=6696000,\ &
\chi_4=6899079264,\\
\chi_5=146325270,\ &
\chi_6=2450240,\ &
\chi_7=30380,\ &
\chi_8=248.
\end{array}
\end{equation}
By substituting these into the $E_8$ curve,
one obtains
\be
y^2=4x^3-u^2x^2+4u^5.
\end{equation}
$f(u),g(u)$ of the Weierstrass form and its discriminant
are given immediately as
\ba
f\Eqn{=}\frac{1}{12}u^4,\\
g\Eqn{=}\frac{1}{216}u^5(u-864),\\[1ex]
\Delta\Eqn{=}u^{10}(u-432).
\ea
According to Table \ref{SingFibers}, one in fact finds 
the degeneration of type $II^\ast$ or $E_8$ at $u=0$.

\noindent
$\bullet\ D_8$\\
As is well-known, $E_8$ representations contain two types of
weights, those with integer components and half-integer components.
They form independent $D_8$ representations.
If we set the parameters at $\vec{m}=(0,0,0,0,0,0,0,2\pi)$,
every integer weight contributes $+1$ while half-integer weight 
contributes $-1$ to the character.
Then the ``twisted'' characters take values
\be
\begin{array}{llll}
\chi_1=35,\ &
\chi_2=50,\ &
\chi_3=-960,\ &
\chi_4=-41888,\\
\chi_5=3094,\ &
\chi_6=832,\ &
\chi_7=-84,\ &
\chi_8=-8.
\end{array}
\end{equation}
Substituting these into the $E_8$ curve, we obtain
\ba
f\Eqn{=}\frac{1}{12}(u-64)^2(u^2+128u-8192),\\
g\Eqn{=}\frac{1}{216}(u-64)^3(u-32)(u^2-640u+28672),\\[1ex]
\Delta\Eqn{=}(u-64)^{10}(u-48).
\ea
One finds a $I_4^\ast$ or $D_8$ singularity at $u=64$. 

\noindent
$\bullet\ E_7\oplus A_1$\\
Let us next set all $m_i$ to $\pi/2$.
Then we obtain
\ba
f\Eqn{=}\frac{1}{12}(u-48)^3(u+144),\\
g\Eqn{=}\frac{1}{216}(u-48)^5(u-624),\\[1ex]
\Delta\Eqn{=}(u-48)^9(u-112)^2.
\ea
There exists a singularity of type $III^\ast$ at $u=48$ and
one of type $I_2$ at $u=112$.
This corresponds to the unbroken symmetry $E_7\oplus A_1$.
In general when we set all $\{m_i\}$ equal
to a common value $\mu$ and $\mu$ is generic,
one finds an $E_7$ symmetry. We have $E_7\oplus A_1$ symmetry
at $\mu=(n+1/2)\pi$ and
$E_8$ symmetry at $\mu=n\pi$. 

$D_8$ and $E_7\oplus A_1$ are maximal regular subalgebras of $E_8$.
Examples of other maximal subalgebras are listed in Appendix C.
It is shown that when the Wilson line parameters
are proportional to the fundamental weights $\vec{\mu}_j$ of $E_8$
\be
\vec{m}\propto \vec{\mu}_j, \quad j=1,2,\ldots,8
\end{equation}
there appears an unbroken symmetry group which is obtained from
the $E_8$ Dynkin diagram by deleting its $j$-th vertex.

Let us now reinstate the $\tau$ dependence of the $\aE_8$ curve 
and discuss a few interesting cases.

\noindent
$\bullet\ D_4\oplus D_4$\\
We set the parameters as
\be
m_1=m_2=0,\ m_3=m_4=\pi,
\ m_5=m_6=\pi+\pi\tau,\ m_7=m_8=\pi\tau.
\end{equation}
Then the $\aE_8$ curve becomes
\be\label{orbifoldcurve}
y^2=4x^3-\frac{E_4}{12}\left(u^2-\frac{4}{q\eta^{24}}\right)^2x
        -\frac{E_6}{216}\left(u^2-\frac{4}{q\eta^{24}}\right)^3.
\end{equation}
Discriminant is given by
\be
\Delta=\eta^{24}
\left(u-\frac{2}{q^{1/2}\eta^{12}}\right)^6
\left(u+\frac{2}{q^{1/2}\eta^{12}}\right)^6.
\end{equation}
We find two $D_4$ singularities at $u=\pm 2/(q^{1/2}\eta^{12})$.
The J-invariant $J=1728f^3/\Delta$ of the curve
is $u$-independent and this describes a
half of the orbifold $T^4/{\cal I}_4$ discussed in \cite{Sen}.

We note the following interesting similarities among 
specializations of the $\aE_8$ curve:
\renewcommand{\labelenumi}{(\alph{enumi})}
\begin{enumerate}
\item $m_1=\cdots=m_8=0$
\be
y^2=4x^3-\frac{1}{12}E_4u^4x-\frac{1}{216}E_6u^6+4u^5,
\end{equation}
\item $m_1=m_2=m_3=m_4=m_5=0,\ m_6=\pi,\ m_7=\pi+\pi\tau,\ m_8=\pi\tau$
\be
y^2=4x^3-\frac{1}{12}E_4u^4x-\frac{1}{216}E_6u^6-\frac{4}{q^{1/2}\eta^{12}}u^4,
\end{equation}
\item $\{m_j\}=\left\{\frac{2\pi}{3}(k+\ell\tau)\right\},
\quad k,\ell=-1,0,1\ \mbox{except}\ (k,\ell)=(0,0)$
\be
y^2=4x^3-\frac{1}{12}E_4u^4x-\frac{1}{216}E_6u^6+\frac{4}{q\eta^{24}}u^3.
\end{equation}
\end{enumerate}
J-invariant $J=1728f^3/(f^3-27g^2)$ of these curves can be
written commonly as
\be
J=\frac{{E_4}^3v^2}{\left(\eta^{24}v^2+E_6v-432\right)}
\end{equation}
with
\be
\mbox{(a)}\ \ v=u,\qquad
\mbox{(b)}\ \ v=-q^{1/2}\eta^{12}u^2,\qquad
\mbox{(c)}\ \ v=q\eta^{24}u^3.
\end{equation}

\section{Partial Fixing of Parameters}

\minisection{Extraction of the 4-dim $SU(2)\ N_f\!=\!4$ Curve
             and $SO(8)$ Triality}

Now let us discuss a new type of reduction of our six-dimensional curve to 
four-dimensional ones which maintains the manifest $SL(2,{\bf Z})$ symmetry.
We claim that when we set four of the Wilson line parameters at half-periods
\be
m_5=0,\ m_6=\pi,\ m_7=\pi+\pi\tau,\ m_8=\pi\tau
\label{massvalue}\end{equation}
we recover the four-dimensional $SU(2)$ Seiberg--Witten curve
with $N_f=4$ flavors. 

Let us first recall the $N_f=4$ curve \cite{SW2},
whose explicit form is given by
\be\label{N_f=4curve}
\ty^2=4[W_1W_2W_3+A(W_1T_1(e_2-e_3)+W_2T_2(e_3-e_1)+W_3T_3(e_1-e_2))-A^2N]
\end{equation}
with
\ba
W_i\Eqn{=}\tx-e_i\tu-{e_i}^2R,\\
A\Eqn{=}(e_1-e_2)(e_2-e_3)(e_3-e_1),\\
R\Eqn{=}\frac{1}{2}\dsum_i{M_i}^2,\\
T_1\Eqn{=}\frac{1}{12}\dsum_{i>j}{M_i}^2{M_j}^2
  -\frac{1}{24}\dsum_i{M_i}^4,\\
T_2\Eqn{=}-\frac{1}{2}\dprod_iM_i
  -\frac{1}{24}\dsum_{i>j}{M_i}^2{M_j}^2+\frac{1}{48}\dsum_i{M_i}^4,\\
T_3\Eqn{=} \frac{1}{2}\dprod_iM_i
  -\frac{1}{24}\dsum_{i>j}{M_i}^2{M_j}^2+\frac{1}{48}\dsum_i{M_i}^4,\\
N\Eqn{=}\frac{3}{16}\dsum_{i>j>k}{M_i}^2{M_j}^2{M_k}^2
  -\frac{1}{96}\dsum_{i\ne j}{M_i}^2{M_j}^4+\frac{1}{96}\dsum_i{M_i}^6
\ea
where $M_1,\ldots,M_4$ denote bare masses of matter hypermultiplets and
\be
e_1=\frac{1}{12}({\varth_3}^4+{\varth_4}^4),\quad
e_2=\frac{1}{12}({\varth_2}^4-{\varth_4}^4),\quad
e_3=\frac{1}{12}(-{\varth_2}^4-{\varth_3}^4).
\end{equation}

We note that when four of the Wilson line parameters $\{m_i\}$ 
are set at zeros of the theta-function (\ref{massvalue}),
all the odd coefficients $a_3,b_1,b_3,b_5$ vanish identically in our curve.
This is due to the structure of Hecke-type transformation
in the coefficient functions
$\{a_i,b_i\}$. See Appendix A.

Then if we redefine $u^2$ as $u$, the $\aE_8$ curve becomes
\be
y^2=4x^3-\left(a_0u^2+a_2u+a_4\right)x 
-\left(b_0u^3+b_2u^2+b_4u+b_6\right).
\label{preD4curve1}\end{equation}
By comparing 
(\ref{N_f=4curve}) with (\ref{preD4curve1}) one finds that
these curves are in fact the same if one makes
the following identifications of four-dimensional masses with
Wilson line parameters
\ba
\label{4dmass_1}
M_1\Eqn{=}
  \left(\prod_{j=1}^4\varth_1(m_j|\tau)-\prod_{j=1}^4\varth_2(m_j|\tau)\right)
       \Biggm/\prod_{j=1}^4\varth_1(m_j|\tau)\, ,\\
M_2\Eqn{=}
  \left(\prod_{j=1}^4\varth_1(m_j|\tau)+\prod_{j=1}^4\varth_2(m_j|\tau)\right)
       \Biggm/\prod_{j=1}^4\varth_1(m_j|\tau)\, ,\\
M_3\Eqn{=}
  \left(\prod_{j=1}^4\varth_3(m_j|\tau)-\prod_{j=1}^4\varth_4(m_j|\tau)\right)
       \Biggm/\prod_{j=1}^4\varth_1(m_j|\tau)\, ,\\
\label{4dmass_4}
M_4\Eqn{=}
  \left(\prod_{j=1}^4\varth_3(m_j|\tau)+\prod_{j=1}^4\varth_4(m_j|\tau)\right)
       \Biggm/\prod_{j=1}^4\varth_1(m_j|\tau)\, .
\ea
(Common denominator of $\{M_i\}$ 
can be chosen arbitrarily.
We have fixed it to $\prod_j\varth_1(m_j|\tau)$ so that
transformation laws of $\{M_i\}$ fit 
with the convention of \cite{SW2}.)
$x,y,u$ of the curve (\ref{preD4curve1})
and $\tilde{x},\tilde{y},\tilde{u}$ of 
the $N_f=4$ curve (\ref{N_f=4curve}) are related as
\ba
u\Eqn{=}L^2\tu-\frac{1}{24\eta^{24}}
  \left(\frac{1}{12}{E_4}^2a_2-E_6b_2\right),\\
x\Eqn{=}L^2\left(\tx-\frac{1}{144}E_4\sum_{i=1}^{4} {M_i}^2\right),\\
y\Eqn{=}L^3\ty
\ea
with
\be
L=\frac{1}{q^{1/4}\eta^{18}}\prod_{j=1}^4\varth_1(m_j|\tau).
\end{equation}

The parameter $\tau$ was interpreted in the $E$-string theory as
the modulus of the torus $T^2$ of fifth and sixth dimensions. 
Now we identify it with the bare coupling of four-dimensional gauge theory.
We have derived the $SL(2,{\bf Z})$ symmetry of
four-dimensional $N_f=4$ theory from the geometry of six dimensions.
This relationship has in fact been suggested before \cite{GMS}.

The structure of four-dimensional masses $\{M_j\}$
(\ref{4dmass_1})--(\ref{4dmass_4}) is quite interesting;
they are invariant under the subgroup $\Gamma(2)$ of the modular group
which consists of matrices of the form
$\biggl(\begin{array}{cc}a&b\\c&d\end{array}\biggr)
\equiv\biggl(\begin{array}{cc}1&0\\0&1\end{array}\biggr) \bmod 2$
with $ad-bc=1$. On the other hand they do transform under $S$ and $T$ 
transformations of the modular group. In fact
the quotient $SL(2,{\bf Z})/\Gamma(2)$ equals the symmetric group 
${\bf S}_3=\{I,S,T,ST,TST,T^{-1}S\}$.
$\{M_j\}$ transform under the action of ${\bf S}_3$ as\\
$\tau\to\tau+1$ :
\ba
M_1\Eqn{\to}M_1,\\
M_2\Eqn{\to}M_2,\\
M_3\Eqn{\to}M_3,\\
M_4\Eqn{\to}-M_4,
\ea
$\tau\to-\dfrac{1}{\tau}$ :
\ba
M_1\Eqn{\to}\frac{1}{2}(M_1+M_2+M_3-M_4),\\
M_2\Eqn{\to}\frac{1}{2}(M_1+M_2-M_3+M_4),\\
M_3\Eqn{\to}\frac{1}{2}(M_1-M_2+M_3+M_4),\\
M_4\Eqn{\to}\frac{1}{2}(-M_1+M_2+M_3+M_4).
\ea
This is exactly the action of $SO(8)$ triality transformation of
the $N_f=4$ theory proposed in \cite{SW2}.
Triality has been postulated for the consistency of physical interpretation of
four-dimensional gauge theory. Here it has been derived naturally from a 
six-dimensional setting.

Note that the four-dimensional masses $M_j(m_i;\tau)$
are proportional to characters of $\hD_4$ Weyl orbits at level one
\ba
w_{\rm b}^{\aD_4}(m_i;\tau)
  \Eqn{=}\frac{1}{2}\left(\prod_{j=1}^4\varth_3(m_j|\tau)
                         +\prod_{j=1}^4\varth_4(m_j|\tau)\right),\\
w_{\rm v}^{\aD_4}(m_i;\tau)
 \Eqn{=}\frac{1}{2}\left(\prod_{j=1}^4\varth_3(m_j|\tau)
                        -\prod_{j=1}^4\varth_4(m_j|\tau)\right),\\
w_{\rm s}^{\aD_4}(m_i;\tau)
 \Eqn{=}\frac{1}{2}\left(\prod_{j=1}^4\varth_2(m_j|\tau)
                        +\prod_{j=1}^4\varth_1(m_j|\tau)\right),\\
w_{\rm c}^{\aD_4}(m_i;\tau)
 \Eqn{=}\frac{1}{2}\left(\prod_{j=1}^4\varth_2(m_j|\tau)
                        -\prod_{j=1}^4\varth_1(m_j|\tau)\right)
\ea
where subscripts $({\rm b,v,s,c})$ refer to
the (basic,vector,spinor,conjugate-spinor)
representations of $SO(8)$, respectively.

\minisection{Donagi--Witten Curve}

Recall that if one sets the four masses $M_i$ of the $N_f=4$ curve at
\be
M_1=M_2=\frac{M}{2},\quad M_3=M_4=0,
\end{equation}
one obtains the Seiberg--Witten curve for the $SU(2)$ theory
with an adjoint matter \cite{SW2}
\be\label{adjcurve}
\ty^2=4\prod_{k=1}^3\left(\tx-e_k\tu-\frac{1}{4}{e_k}^2M^2\right).
\end{equation}
This is in fact the curve discussed by Donagi and Witten \cite{DW}.
To obtain this curve directly from the $\aE_8$ curve,
one may set
\be
m_2=0,\ m_3=m_4=\pi,
\ m_5=m_6=\pi+\pi\tau,\ m_7=m_8=\pi\tau
\end{equation}
and transform the variables as
\ba
u^2\Eqn{=}L^2\Bigl(\tu-\wp(m_1)\Bigr),\\
x\Eqn{=}L^2\left(\tx-\frac{1}{72}E_4\right),\\
y\Eqn{=}L^3\ty
\ea
with
\be
L=\frac{2i}{q^{1/2}\eta^{15}}\varth_1(m_j|\tau).
\end{equation}
Then one obtains a curve
\be
\ty^2=4\prod_{k=1}^3\left(\tx-e_k\tu-{e_k}^2\right),
\end{equation}
which agrees with (\ref{adjcurve}) after a rescaling.

A characteristic feature of the curve (\ref{adjcurve}) is
the factored form of the cubic in $x$.
This form ensures that each singularity of the $u$-plane
is at least doubled 
since the discriminant of a curve of the form
\be
y^2=4(x-\alpha)(x-\beta)(x-\gamma)
\end{equation}
is given by
\be
\Delta=16(\alpha-\beta)^2(\beta-\gamma)^2(\gamma-\alpha)^2.
\end{equation}

A more general factorized curve may be obtained from
the $\aE_8$ curve.
Let us leave $m_1,m_2$ free and fix the others at
\ba
m_3=m_4=\pi,\ m_5=m_6=\pi+\pi\tau,\ m_7=m_8=\pi\tau.
\ea
Then one obtains a curve
\be
y^2=4\prod_{k=1}^3
  \biggl(x-\left(x_0^{(k)}u^2+x_1^{(k)}u+x_2^{(k)}\right)\biggr)
\end{equation}
where
\be
x_0^{(k)}=e_k,
\end{equation}
\ba
x_1^{(1)}\Eqn{=}
\frac{2{\varth_2}^4\varth_1(m_1|\tau)\varth_1(m_2|\tau)}{q^{1/2}\eta^6E_4},\\
x_1^{(2)}\Eqn{=}
\frac{2(-{\varth_3}^4)\varth_1(m_1|\tau)\varth_1(m_2|\tau)}
     {q^{1/2}\eta^6E_4},\\
x_1^{(3)}\Eqn{=}
\frac{2{\varth_4}^4\varth_1(m_1|\tau)\varth_1(m_2|\tau)}{q^{1/2}\eta^6E_4},
\ea
\ba
x_2^{(k)}\Eqn{=}
\frac{\varth_1(m_1|\tau)^2\varth_1(m_2|\tau)^2}{q\,\eta^{36}}
\Biggl[-4e_k\wp(m_1)\wp(m_2)\no\\\Eqn{}
  +\left(\frac{1}{18}E_4-4{e_k}^2\right)
    \Bigl(\wp(m_1)+\wp(m_2)\Bigr)
  +E_6\left(\frac{1}{108}-\frac{7}{6}\frac{{e_k}^2}{E_4}
   +24\frac{{e_k}^4}{{E_4}^2}\right)
\Biggr].\no\\
\ea
This curve describes
six $I_2$ fibers and thus corresponds to
$A_1^{\oplus 6}$ symmetry.

\section{Discussions}

In this paper we have discussed various properties of
the Seiberg--Witten curve for the $E$-string theory
which we obtained previously. First we have studied its holomorphic
sections and identified the geometrical meaning of the Wilson line parameters
$\{m_i\}$. By fine tuning these parameters we have then studied
various degenerations 
of the curve which correspond to several unbroken symmetry groups.

We have also presented a new way of reduction to four dimensions
without taking the
degenerate limit of $T^2$ so that the $SL(2,{\bf Z})$ symmetry is left intact. 
By setting some of the parameters $\{m_i\}$ to special values
we have obtained the
four-dimensional Seiberg--Witten theory with $N_f=4$ flavors and also the
curve by Donagi and Witten describing a perturbed ${\cal N}=4$ theory.
In this reduction four-dimensional masses are expressed in terms of
theta functions and possess exactly the proposed triality properties.
Thus our curve seems to serve successfully as a geometrical way
of deriving $SL(2,{\bf Z})$ symmetry from higher dimensions.

We should point out, however, we have redefined
$u^2$ as $u$ in deriving four-dimensional curves in the new reduction.
The geometrical significance of this change of variable is
yet to be understood.
We hope that we will be able to discuss this issue in a future publication.

\subsection*{Acknowledgments}
Authors would like to thank Y.~Yamada and Y.~Shimizu for helpful comments.
K.S. would like to thank Y.~Hikida and S.~Yamaguchi
for valuable discussions.

\renewcommand{\thesubsection}{Appendix
  \Alph{subsection}.\hspace{-.5em}}
\makeatletter
\@addtoreset{equation}{subsection}
\renewcommand{\theequation}{\Alph{subsection}.\arabic{equation}}
\makeatother
\setcounter{subsection}{0}

\section{$\hE_8$ Curve}

Curve:
\ba
y^2&=&4x^3-\left(a_0u^4+a_1u^3+a_2u^2+a_3u+a_4\right)x \no\\\Eqn{}
-\left(b_0u^6+b_1u^5+b_2u^4+b_3u^3+b_4u^2+b_5u+b_6\right).
\ea

\noindent
Coefficients:
\ba
a_0\Eqn{=}\frac{1}{12}E_4,\
b_0=\frac{1}{216}E_6,\
a_1=0,\
b_1=-4{P(m_i;\tau)\over E_4},\\
a_2\Eqn{=}\frac{1}{E_4\Delta}\biggl[
f_{a2,0}(\tau) P(2 m_i;2\tau)
+f_{a2,1}(\tau) P\Bigl(m_i;\frac{\tau}{2}\Bigr)
\no\\ \Eqn{}+f_{a2,1}(\tau+1)
P\Bigl(m_i;\frac{\tau+1}{2}\Bigr)\biggr]\ea
where\ba
f_{a2,0}(\tau)\Eqn{=}\frac{2}{3}\left(E_4(\tau)-9\varth_2(\tau)^8\right),\\
f_{a2,1}(\tau)\Eqn{=}\frac{1}{24}\left(E_4(\tau)-9\varth_4(\tau)^8\right),\ea \ba
&&b_2=\frac{1}{{E_4}^2\Delta}\biggl[
f_{b2,0}(\tau) P(2 m_i;2\tau)
+f_{b2,1}(\tau) P\Bigl(m_i;\frac{\tau}{2}\Bigr)
\no\\&&\hspace{2em} +f_{b2,1}(\tau+1) P\Bigl(m_i;\frac{\tau+1}{2}\Bigr)\biggr]\ea
where\ba f_{b2,0}(\tau)&=&
  \frac{1}{36}\left(\varth_3(\tau)^4+\varth_4(\tau)^4\right)
  \left(E_4(\tau)^2+60 E_4(\tau)\varth_2(\tau)^8\right.
 \no\\ \Eqn{} \left.-45\varth_2(\tau)^{16}\right), \\
f_{b2,1}(\tau)&=&
  -\frac{1}{576}\left(\varth_3(\tau)^4+\varth_2(\tau)^4\right)
  \left(E_4(\tau)^2+60 E_4(\tau)\varth_4(\tau)^8\right.
  \no\\ \Eqn{}\left.-45\varth_4(\tau)^{16}\right),\ea
\ba
a_3\Eqn{=}\frac{1}{{E_4}^2\Delta^2}\biggl[
f_{a3,0}(\tau) P(3 m_i;3\tau)
+f_{a3,1}(\tau) P\Bigl(m_i;\frac{\tau}{3}\Bigr)
\no\\\Eqn{} +f_{a3,1}(\tau+1) P\Bigl(m_i;\frac{\tau+1}{3}\Bigr)
+f_{a3,1}(\tau+2) P\Bigl(m_i;\frac{\tau+2}{3}\Bigr)
\no\\ \Eqn{}+\frac{2}{3}E_6(\tau) P(m_i;\tau)^3\biggr]\ea
where\ba
f_{a3,0}(\tau)\Eqn{=}\frac{1}{3}E_4(\tau) h_2(\tau)^2
  \left(7 E_4(\tau)-9 h_0(\tau)^4\right),\\
f_{a3,1}(\tau)\Eqn{=}-\frac{1}{3^8}E_4(\tau) h_3\Bigl(\frac{\tau}{3}\Bigr)^2
  \left(7 E_4(\tau)-h_0\Bigl(\frac{\tau}{3}\Bigr)^4\right),\ea
\ba
b_3\Eqn{=}\frac{1}{{E_4}^3\Delta^2}\biggl[
f_{b3,0}(\tau) P(3 m_i;3\tau)
+f_{b3,1}(\tau) P\Bigl(m_i;\frac{\tau}{3}\Bigr)
\no\\\Eqn{} +f_{b3,1}(\tau+1) P\Bigl(m_i;\frac{\tau+1}{3}\Bigr)
+f_{b3,1}(\tau+2) P\Bigl(m_i;\frac{\tau+2}{3}\Bigr)
\no\\\Eqn{}+\frac{1}{54}\left(8{E_4}^3-5{E_6}^2\right) P(m_i;\tau)^3\biggr]\ea
where\ba
f_{b3,0}(\tau)\Eqn{=}\frac{1}{18}E_4(\tau)^2 h_2(\tau)^2
  \left(32 h_2(\tau)^2+48h_2(\tau) h_0(\tau)^3-81
  h_0(\tau)^6\right),\no \\ \Eqn{}\\
f_{b3,1}(\tau)\Eqn{=}\frac{1}{2\cdot 3^{12}} E_4(\tau)^2 h_3
  \Bigl(\frac{\tau}{3}\Bigr)^2
  \left(32 h_3\Bigl(\frac{\tau}{3}\Bigr)^2 +48
  h_3\Bigl(\frac{\tau}{3}\Bigr)
  h_0\Bigl(\frac{\tau}{3}\Bigr)^3\right.
  \no\\\Eqn{}\left.-81h_0\Bigl(\frac{\tau}{3}\Bigr)^6\right),\ea
\ba a_4\Eqn{=}
\frac{1}{E_4\Delta^2} \biggl[
 16            \varth_4(2\tau)^8 P(4m_i;4\tau)
+\frac{1}{256} \varth_2\Bigl(\frac{\tau}{2}\Bigr)^8
  P\Bigl(m_i;\frac{\tau}{4}\Bigr)\no\\
&&\hspace{10mm}
+\frac{1}{256} \varth_2\Bigl(\frac{\tau+1}{2}\Bigr)^8
  P\Bigl(m_i;\frac{\tau+1}{4}\Bigr)
+\frac{1}{256} \varth_2\Bigl(\frac{\tau+2}{2}\Bigr)^8
  P\Bigl(m_i;\frac{\tau+2}{4}\Bigr)\no\\
&&\hspace{10mm}
+\frac{1}{256} \varth_2\Bigl(\frac{\tau+3}{2}\Bigr)^8
   P\Bigl(m_i;\frac{\tau+3}{4}\Bigr)
+              \varth_4(2\tau)^8
   P\Bigl(2m_i;\tau+\frac{1}{2}\Bigr)\no\\
&&\hspace{10mm}
+E_4 P(2m_i;\tau)
-\frac{3}{2}\Delta^2{a_2}^2
+\frac{3}{8} E_4\Delta a_2 {b_1}^2
-\frac{9}{128} {E_4}^2{b_1}^4\biggr],\\
\no\\
b_4\Eqn{=}
\frac{1}{E_4\Delta^2}\biggl[
f_{b4,0}(\tau) P(4 m_i;4\tau)
+f_{b4,1}(\tau) P\Bigl(m_i;\frac{\tau}{4}\Bigr)\no\\
&&\hspace{10mm}
+f_{b4,1}(\tau+1) P\Bigl(m_i;\frac{\tau+1}{4}\Bigr)
+f_{b4,1}(\tau+2) P\Bigl(m_i;\frac{\tau+2}{4}\Bigr)\no\\
&&\hspace{10mm}
+f_{b4,1}(\tau+3) P\Bigl(m_i;\frac{\tau+3}{4}\Bigr)
+f_{b4,2}(\tau) P\Bigl(2 m_i;\tau+\frac{1}{2}\Bigr)\no\\
&&\hspace{10mm}
+\frac{5}{48}E_6 P(2 m_i;\tau)
-\frac{1}{96}E_4 E_6\Delta b_3 b_1
-3 E_4\Delta b_2 {b_1}^2
-\frac{55}{384}E_4 E_6 {b_1}^4
\biggr]\no\\ \Eqn{}\ea
where \ba
f_{b4,0}(\tau)\Eqn{=}-\frac{8}{9}\Bigl(32\varth_3(2\tau)^{12}
  -75\varth_3(2\tau)^4\varth_4(2\tau)^8+2\varth_4(2\tau)^{12}\Bigr),\\
f_{b4,1}(\tau)\Eqn{=}\frac{1}{2^{11}\cdot 9}\left(32\varth_3
  \Bigl(\frac{\tau}{2}\Bigr)^{12}
  -75\varth_3\Bigl(\frac{\tau}{2}\Bigr)^4
  \varth_2\Bigl(\frac{\tau}{2}\Bigr)^8
  +2\varth_2\Bigl(\frac{\tau}{2}\Bigr)^{12}\right),\\
f_{b4,2}(\tau)\Eqn{=}-\frac{1}{18}\Bigl(32\varth_2(2\tau)^{12}
  -75\varth_2(2\tau)^4\varth_4(2\tau)^8-2\varth_4(2\tau)^{12}\Bigr),\ea
\ba
b_5\Eqn{=}
\frac{1}{{E_4}^2\Delta^3}\biggl[
f_{b5,0}(\tau) P(5 m_i;5\tau)
+f_{b5,1}(\tau) P\Bigl(m_i;\frac{\tau}{5}\Bigr)
+f_{b5,1}(\tau+1) P\Bigl(m_i;\frac{\tau+1}{5}\Bigr)\no\\
&&\hspace{12mm}
+f_{b5,1}(\tau+2) P\Bigl(m_i;\frac{\tau+2}{5}\Bigr)
+f_{b5,1}(\tau+3) P\Bigl(m_i;\frac{\tau+3}{5}\Bigr)\no\\
&&\hspace{12mm}
+f_{b5,1}(\tau+4) P\Bigl(m_i;\frac{\tau+4}{5}\Bigr)\no\\
&&\hspace{8mm}
+\biggl(
g_{b5,0}(\tau) P(4 m_i;4\tau)
+g_{b5,1}(\tau) P\Bigl(m_i;\frac{\tau}{4}\Bigr)\no\\
&&\hspace{12mm}
+g_{b5,1}(\tau+1) P\Bigl(m_i;\frac{\tau+1}{4}\Bigr)
+g_{b5,1}(\tau+2) P\Bigl(m_i;\frac{\tau+2}{4}\Bigr)\no\\
&&\hspace{12mm}
+g_{b5,1}(\tau+3) P\Bigl(m_i;\frac{\tau+3}{4}\Bigr)
+g_{b5,2}(\tau) P\Bigl(2 m_i;\tau+\frac{1}{2}\Bigr)\biggr)P(m_i;\tau)\biggr]
\no\\
&&
+\frac{1}{\Delta^3}\biggl[
 \frac{3}{2}\Delta^2 b_3{b_1}^2
+\frac{3}{16}E_6\Delta b_2{b_1}^3
+\frac{1}{9216}(53{E_4}^3+67{E_6}^2){b_1}^5
\biggr]\ea
where
\ba
&&f_{b5,0}(\tau)=\frac{2}{3}\eta(\tau)^{16}\eta(5\tau)^{16}\\
&&\hspace{14mm}\times
\biggl(5^{11}+4\cdot 5^9\frac{\eta(\tau)^{6}}{\eta(5\tau)^{6}}
             +22\cdot 5^6\frac{\eta(\tau)^{12}}{\eta(5\tau)^{12}}
             +4\cdot 5^4\frac{\eta(\tau)^{18}}{\eta(5\tau)^{18}}
             -31
	     \frac{\eta(\tau)^{24}}{\eta(5\tau)^{24}}\biggr),\no \\
&&f_{b5,1}(\tau)=\frac{2}{3}\eta(\tau)^{16}\eta(\tfrac{\tau}{5})^{16}\no\\
&&\hspace{14mm}\times
\biggl(\frac{1}{5}+4\frac{\eta(\tau)^{6}}{\eta(\frac{\tau}{5})^{6}}
             +22\frac{\eta(\tau)^{12}}{\eta(\frac{\tau}{5})^{12}}
             +20\frac{\eta(\tau)^{18}}{\eta(\frac{\tau}{5})^{18}}
             -31\frac{\eta(\tau)^{24}}{\eta(\frac{\tau}{5})^{24}}\biggr),\\
&&g_{b5,0}(\tau)=
\frac{2}{3}\varth_3(2\tau)^4\varth_4(2\tau)^8
  \Bigl(-1024\varth_3(2\tau)^{12}
        +1664\varth_3(2\tau)^8\varth_4(2\tau)^4\no\\
&&\hspace{40mm}
        - 644\varth_3(2\tau)^4\varth_4(2\tau)^8
        +  55\varth_4(2\tau)^{12}\Bigr),\\
&&g_{b5,1}(\tau)=
\frac{1}{2^{19}\cdot 3}\varth_3\Bigl(\frac{\tau}{2}\Bigr)^4
  \varth_2\Bigl(\frac{\tau}{2}\Bigr)^8
  \biggl(-1024\varth_3\Bigl(\frac{\tau}{2}\Bigr)^{12}
        +1664\varth_3\Bigl(\frac{\tau}{2}\Bigr)^8
             \varth_2\Bigl(\frac{\tau}{2}\Bigr)^4\no\\
&&\hspace{40mm}
        - 644\varth_3\Bigl(\frac{\tau}{2}\Bigr)^4
             \varth_2\Bigl(\frac{\tau}{2}\Bigr)^8
        +  55\varth_2\Bigl(\frac{\tau}{2}\Bigr)^{12}\biggr),\\
&&g_{b5,2}(\tau)=
-\frac{1}{24}\varth_2(2\tau)^4\varth_4(2\tau)^8
 \Bigl( 1024\varth_2(2\tau)^{12}
       +1664\varth_2(2\tau)^8\varth_4(2\tau)^4\no\\
&&\hspace{40mm}
       + 644\varth_2(2\tau)^4\varth_4(2\tau)^8
       +  55\varth_4(2\tau)^{12}\Bigr),\ea
\ba
b_6\Eqn{=}
\frac{1}{E_4\Delta^3}\biggl[
f_{b6,0}(\tau) P(6 m_i;6\tau)
+f_{b6,1}(\tau) P\Bigl(m_i;\frac{\tau}{6}\Bigr)
\no\\&&\hspace{6mm}+f_{b6,1}(\tau+1) P\Bigl(m_i;\frac{\tau+1}{6}\Bigr)
+f_{b6,1}(\tau+2) P\Bigl(m_i;\frac{\tau+2}{6}\Bigr)
\no\\&&\hspace{10mm}+f_{b6,1}(\tau+3) P\Bigl(m_i;\frac{\tau+3}{6}\Bigr)
+f_{b6,1}(\tau+4) P\Bigl(m_i;\frac{\tau+4}{6}\Bigr)
\no\\
&&\hspace{6mm}
+f_{b6,1}(\tau+5) P\Bigl(m_i;\frac{\tau+5}{6}\Bigr)
+f_{b6,2}(\tau) P\Bigl(3 m_i;\frac{3\tau}{2}\Bigr)
\no\\
&&\hspace{6mm}+f_{b6,2}(\tau+1) P\Bigl(3 m_i;\frac{3\tau+1}{2}\Bigr)
+f_{b6,3}(\tau) P\Bigl(2 m_i;\frac{2\tau}{3}\Bigr)\no\\
&&\hspace{6mm}
+f_{b6,3}(\tau+1) P\Bigl(2 m_i;\frac{2\tau+2}{3}\Bigr)
+f_{b6,3}(\tau+2) P\Bigl(2 m_i;\frac{2\tau+1}{3}\Bigr)\biggr]\no\\
\Eqn{+}\frac{a_2}{E_4\Delta^2}\biggl[
g_{b6,0}(\tau) P(4 m_i;4\tau)
+g_{b6,1}(\tau) P\Bigl(m_i;\frac{\tau}{4}\Bigr)\no\\
&&\hspace{6mm}
+g_{b6,1}(\tau+1) P\Bigl(m_i;\frac{\tau+1}{4}\Bigr)
+g_{b6,1}(\tau+2) P\Bigl(m_i;\frac{\tau+2}{4}\Bigr)\no\\
&&\hspace{6mm}
+g_{b6,1}(\tau+3) P\Bigl(m_i;\frac{\tau+3}{4}\Bigr)
+g_{b6,2}(\tau) P\Bigl(2 m_i;\tau+\frac{1}{2}\Bigr)\biggr]\no\\
\Eqn{+}\frac{1}{\Delta^3}\biggl[
  -\frac{83}{7344} {E_6}{\Delta}^2b_5{b_1}
  +\frac{83}{408} {\Delta}^2b_4{b_1}^2
  +\frac{1}{34} {\Delta}^2{b_3}{b_2}{b_1}\no\\
&&\hspace{5mm}
  +\frac{29}{6528} {E_6}{\Delta}{b_3}{b_1}^3
  +\frac{669}{1088} {\Delta}{b_2}{b_1}^4
  -\frac{5}{7344} {E_4}{\Delta}^2{a_3}{a_2}{b_1}\no\\
&&\hspace{5mm}
  -\frac{419}{235008} {E_4}^2{\Delta}{a_3}{b_1}^3
  -\frac{1}{36864} {E_4}{E_6}{\Delta}{a_2}^2{b_1}^2
  +\frac{1}{1536} {E_4}^2{\Delta}{a_2}{b_2}{b_1}^2\no\\
&&\hspace{5mm}
  -\frac{1}{256} {E_6}{\Delta}{b_2}^2{b_1}^2
  +\frac{1215}{69632} {E_6}{b_1}^6\biggr]\ea
where
\ba
&&f_{b6,0}(\tau)=
-\frac{4}{17}\Bigl(h_0(\tau)+h_0(2\tau)\Bigr)
\Bigl(h_0(\tau)-2 h_0(2\tau)\Bigr)^2\\
&&\hspace{1mm}\times
\Bigl(27h_0(\tau)^3+84h_0(\tau)^2 h_0(2\tau)
+72 h_0(\tau)h_0(2\tau)^2-32h_0(2\tau)^3\Bigr),\no\\
&&f_{b6,1}(\tau)=
\frac{1}{297432}\left(2 h_0\Bigl(\frac{\tau}{3}\Bigr)
  +h_0\Bigl(\frac{\tau}{6}\Bigr)\right)
\left(h_0\Bigl(\frac{\tau}{3}\Bigr)
-h_0\Bigl(\frac{\tau}{6}\Bigr)\right)^2\\
&&\hspace{1mm}\times
\left(27h_0\Bigl(\frac{\tau}{3}\Bigr)^3
  +42h_0\Bigl(\frac{\tau}{3}\Bigr)^2 h_0\Bigl(\frac{\tau}{6}\Bigr)
+18 h_0\Bigl(\frac{\tau}{3}\Bigr)h_0\Bigl(\frac{\tau}{6}\Bigr)^2
-4h_0\Bigl(\frac{\tau}{6}\Bigr)^3\right),\no\\
&&f_{b6,2}(\tau)=
\frac{1}{136}\left(-2 h_0(\tau)+h_0\Bigl(\frac{\tau}{2}\Bigr)\right)
\left(h_0(\tau)+h_0\Bigl(\frac{\tau}{2}\Bigr)\right)^2\\
&&\hspace{1mm}\times
\left(27h_0(\tau)^3-42h_0(\tau)^2 h_0\Bigl(\frac{\tau}{2}\Bigr)
+18 h_0(\tau)h_0\Bigl(\frac{\tau}{2}\Bigr)^2
+4h_0\Bigl(\frac{\tau}{2}\Bigr)^3\right)\no\ea
\ba
&&f_{b6,3}(\tau)=
\frac{4}{37179}\left(h_0\Bigl(\frac{\tau}{3}\Bigr)
-h_0\Bigl(\frac{2\tau}{3}\Bigr)\right)
\left(h_0\Bigl(\frac{\tau}{3}\Bigr)
+2 h_0\Bigl(\frac{2\tau}{3}\Bigr)\right)^2\\
&&\hspace{1mm}\times
\left(27h_0\Bigl(\frac{\tau}{3}\Bigr)^3
-84h_0\Bigl(\frac{\tau}{3}\Bigr)^2 h_0\Bigl(\frac{2\tau}{3}\Bigr)
+72 h_0\Bigl(\frac{\tau}{3}\Bigr)h_0\Bigl(\frac{2\tau}{3}\Bigr)^2
+32h_0\Bigl(\frac{2\tau}{3}\Bigr)^3\right),\no
\ea
\ba
&&g_{b6,0}(\tau)=
-\frac{640}{51} \varth_3(2\tau)^4+\frac{32}{9}\varth_4(2\tau)^4,\\
&&g_{b6,1}(\tau)=
\frac{5}{408}\varth_3\Bigl(\frac{\tau}{2}\Bigr)^4
-\frac{1}{288}\varth_2\Bigl(\frac{\tau}{2}\Bigr)^4,\\
&&g_{b6,2}(\tau)=
-\frac{40}{51}\varth_2(2\tau)^4-\frac{2}{9} \varth_4(2\tau)^4.
\ea\\
Here
\ba
P(m_i;\tau)\Eqn{=}\frac{1}{2}\sum_{\ell=1}^4\prod_{j=1}^8\varth_\ell(m_j|\tau)
\ea
is the $\aE_8$ Weyl orbit character at level one.
Theta functions are defined as
\ba
\varth_1(z|\tau)\Eqn{=}
  i\sum_{n\in \bf Z}(-1)^n y^{n-1/2}q^{(n-1/2)^2/2},\\
\varth_2(z|\tau)\Eqn{=}
  \sum_{n\in \bf Z}y^{n-1/2}q^{(n-1/2)^2/2},\\
\varth_3(z|\tau)\Eqn{=}
  \sum_{n\in \bf Z}y^n q^{n^2/2},\\
\varth_4(z|\tau)\Eqn{=}
  \sum_{n\in \bf Z}(-1)^n y^n q^{n^2/2}
\ea
where $y=e^{i z},\ q=e^{2\pi i \tau}$
and we abbreviate $\varth_\ell(0|\tau)$
to $\varth_\ell(\tau)$ or just $\varth_\ell$.\\
$E_{2n}$ is the Eisenstein series with weight $2n$,
\be
E_{2n}(\tau)=1+{(2\pi i)^{2n}\over (2n-1)!\,\zeta(2n)}
\sum_{m=1}^{\infty}\frac{m^{2n-1}q^m}{1-q^m},
\end{equation}
and
$\Delta(\tau)\equiv\eta(\tau)^{24}
  =\frac{1}{1728}\Bigl({E_4(\tau)}^3-{E_6(\tau)}^2\Bigr)$.
Functions $\{h_i\}$ are defined by
\ba
h_0(\tau)\Eqn{=}\sum_{n_1,n_2=-\infty}^{\infty}q^{n_1^2+n_2^2-n_1n_2}
=\varth_3(2\tau)\varth_3(6\tau)+\varth_2(2\tau)\varth_2(6\tau),\\
h_2(\tau)\Eqn{=}\frac{\eta(\tau)^9}{\eta(3\tau)^3},\quad
h_3(\tau)=27\frac{\eta(3\tau)^9}{\eta(\tau)^3}.
\ea

\section{$E_n$ Curves and Holomorphic Sections}

\noindent
$E_8$ curve:
\begin{eqnarray}
y^2 \Eqn{=} 4x^3\no\\
&&+\Bigl(-u^2+4\chi_1-100\chi_8+9300\Bigr)x^2\no\\
&&+\Bigl( (2\chi_2-12\chi_7-70\chi_1+1840\chi_8-115010)u\no\\
&&\hspace{1em}+(4\chi_3-4\chi_6-64\chi_1\chi_8+824\chi_8\chi_8-112\chi_2\no\\
&&\hspace{2em}+680\chi_7+8024\chi_1-205744\chi_8+9606776) \Bigr)x\no\\
&&+\Bigl(4u^5\no\\
&&\hspace{1em}+(4\chi_8-992)u^4\no\\
&&\hspace{1em}+(4\chi_7-12\chi_1-680\chi_8+93620)u^3\no\\
&&\hspace{1em}+(4\chi_6-8\chi_1\chi_8+92\chi_8\chi_8-28\chi_2
              -540\chi_7+2320\chi_1+30608\chi_8\no\\
&&\hspace{2em}-3823912)u^2
            +(4\chi_5-4\chi_1\chi_7-20\chi_2\chi_8+116\chi_7\chi_8
              +8\chi_1\chi_1\no\\
&&\hspace{2em}-52\chi_3-416\chi_6+1436\chi_1\chi_8
-17776\chi_8\chi_8+4180\chi_2+16580\chi_7\no\\
&&\hspace{2em}-182832\chi_1
              +1103956\chi_8+18130536)u\no\\
&&\hspace{1em}+(4\chi_4-\chi_2\chi_2+4\chi_1\chi_1\chi_8-40\chi_3\chi_8
              +36\chi_6\chi_8+248\chi_1\chi_8\chi_8\no\\
&&\hspace{2em}-2232\chi_8\chi_8\chi_8
              +2\chi_1\chi_2-232\chi_5+224\chi_1\chi_7+1124\chi_2\chi_8
              \no\\
&&\hspace{2em}-6580\chi_7\chi_8-457\chi_1\chi_1+4980\chi_3
+8708\chi_6-88136\chi_1\chi_8\no\\
&&\hspace{2em}+1129964\chi_8\chi_8-146282\chi_2
              +66612\chi_7+6123126\chi_1\no\\
&&\hspace{2em}-104097420\chi_8+2630318907)\Bigr)
\end{eqnarray}
where $\chi_1,\ldots,\chi_8$ denote
characters of $E_8$ fundamental representations and are defined as
\ba
&&
\chi_1=\chEei{1}{0}{0}{0}{0}{0}{0}{0}, \ \
\chi_2=\chEei{0}{1}{0}{0}{0}{0}{0}{0}, \ \
\chi_3=\chEei{0}{0}{1}{0}{0}{0}{0}{0},\no\\
&& 
\chi_4=\chEei{0}{0}{0}{1}{0}{0}{0}{0},\ \
\chi_5=\chEei{0}{0}{0}{0}{1}{0}{0}{0}, \ \
\chi_6=\chEei{0}{0}{0}{0}{0}{1}{0}{0}, \no\\
&&
\chi_7=\chEei{0}{0}{0}{0}{0}{0}{1}{0}, \ \
\chi_8=\chEei{0}{0}{0}{0}{0}{0}{0}{1}.
\ea

\noindent
$240$ sections for $E_8$ curve:
\begin{eqnarray}
x\Eqn{=} \frac{1}{4\sin^2\lambda}\times\no\\
&&\Bigl[u^2+\Bigl(2\cos 4\lambda-24 \cos 2\lambda
  +2\chi_7^{E_7} \cos\lambda-90\Bigr) u\no\\
&&\ 
  +\Bigl(2\cos 6\lambda+(2\chi_1^{E_7}-134) \cos 4\lambda\no\\
&&\hspace{1.63em}
  -16 \chi_7^{E_7} \cos 3\lambda
  +(-20 \chi_1^{E_7}+2\chi_6^{E_7}+2396) \cos 2\lambda\no\\
&&\hspace{1.63em}
  -96 \chi_7^{E_7} \cos\lambda
  +\chi_7^{E_7}\chi_7^{E_7}+18 \chi_1^{E_7}
  -2 \chi_6^{E_7}+872\Bigr)\Bigr],\\
y\Eqn{=} \frac{1}{4 \sin^3\lambda}\times\no\\
&&\Bigl[(\cos\lambda) u^3
  +\Bigl(-\cos 5\lambda+5 \cos 3\lambda+\chi_7^{E_7} \cos 2\lambda
  -172 \cos\lambda+2 \chi_7^{E_7}\Bigr) u^2\no\\
&&\ 
  +\Bigl(-3 \cos 7\lambda+(-\chi_1^{E_7}+124) \cos 5\lambda
  +(5 \chi_1^{E_7}+\chi_6^{E_7}-575) \cos 3\lambda\no\\
&&\hspace{1.63em}
  +(-2 \chi_2^{E_7}-108 \chi_7^{E_7}) \cos 2\lambda
  +(3 \chi_7^{E_7}\chi_7^{E_7}-4 \chi_1^{E_7}
  -\chi_6^{E_7}+9862)\cos\lambda\no\\
&&\hspace{1.63em}
  +2 \chi_2^{E_7}-228 \chi_7^{E_7}\Bigr) u\no\\
&&\ 
  +\Bigl(-2 \cos 9\lambda+(-2 \chi_1^{E_7}+176) \cos 7\lambda
  +\chi_2^{E_7} \cos 6\lambda\no\\
&&\hspace{1.63em}
  +(62 \chi_1^{E_7}-\chi_3^{E_7}-\chi_6^{E_7}-3822) \cos 5\lambda\no\\
&&\hspace{1.63em}
  +(\chi_1^{E_7} \chi_7^{E_7}-3 \chi_2^{E_7}
  +\chi_5^{E_7}+3 \chi_7^{E_7}) \cos 4\lambda\no\\
&&\hspace{1.63em}
  +(-\chi_2^{E_7} \chi_7^{E_7}-286 \chi_1^{E_7}+3 \chi_3^{E_7}
  -53 \chi_6^{E_7}+16534) \cos 3\lambda\no\\
&&\hspace{1.63em}
  +(2 \chi_6^{E_7} \chi_7^{E_7}+115 \chi_2^{E_7}-4 \chi_5^{E_7}
  +2906 \chi_7^{E_7}) \cos 2\lambda\no\\
&&\hspace{1.63em}
  +(\chi_2^{E_7} \chi_7^{E_7}-168 \chi_7^{E_7}\chi_7^{E_7}
  +226 \chi_1^{E_7}-2 \chi_3^{E_7}+54 \chi_6^{E_7}-188502)\cos\lambda\no\\
&&\hspace{1.63em}
  +\chi_7^{E_7}\chi_7^{E_7}\chi_7^{E_7}-\chi_1^{E_7} \chi_7^{E_7}
  -2 \chi_6^{E_7} \chi_7^{E_7}-113 \chi_2^{E_7}+3 \chi_5^{E_7}
  +6499 \chi_7^{E_7}\Bigr)\Bigr].\no\\
\end{eqnarray}
where $\chi_1^{E_7},\ldots,\chi_7^{E_7}$ denote
$E_7$ fundamental representation characters and are defined as
\ba
&&
\chi_1^{E_7}=\chEse{1}{0}{0}{0}{0}{0}{0}, \ \
\chi_2^{E_7}=\chEse{0}{1}{0}{0}{0}{0}{0}, \ \
\chi_3^{E_7}=\chEse{0}{0}{1}{0}{0}{0}{0}, \ \
\chi_4^{E_7}=\chEse{0}{0}{0}{1}{0}{0}{0},\no\\
&&
\chi_5^{E_7}=\chEse{0}{0}{0}{0}{1}{0}{0}, \ \
\chi_6^{E_7}=\chEse{0}{0}{0}{0}{0}{1}{0}, \ \
\chi_7^{E_7}=\chEse{0}{0}{0}{0}{0}{0}{1}.
\ea

\vspace{1ex}
\noindent
$E_7$ curve:
\begin{eqnarray}
y^2 \Eqn{=} 4x^3
  +\Bigl(-u^2+4\chi_1^{E_7}-100\Bigr)x^2\\
&&
  +\Bigl( (2\chi_2^{E_7}-12\chi_7^{E_7})u
  + (4\chi_3^{E_7}-4\chi_6^{E_7}-64\chi_1^{E_7}+824) \Bigr)x\no\\
&&
  +\Bigl(4u^4+4\chi_7^{E_7}u^3+(4\chi_6^{E_7}-8\chi_1^{E_7}+92)u^2\no\\
&&\hspace{1.5em}
  +(4\chi_5^{E_7}-4\chi_1^{E_7}\chi_7^{E_7}
    -20\chi_2^{E_7}+116\chi_7^{E_7})u\no\\
&&\hspace{1.5em}
  +(4\chi_4^{E_7}-\chi_2^{E_7}\chi_2^{E_7}+4\chi_1^{E_7}\chi_1^{E_7}
  -40\chi_3^{E_7}+36\chi_6^{E_7}+248\chi_1^{E_7}-2232)\Bigr).\no
\end{eqnarray}

\noindent
$56$ sections for $E_7$ curve:
\begin{eqnarray}
x\Eqn{=}-\Bigl(\Lambda^3+\Lambda^{-3}\Bigr)u
  -\Bigl(\Lambda^6+\chi_1^{E_6} \Lambda^2-8
  +\chi_6^{E_6} \Lambda^{-2}+\Lambda^{-6}\Bigr),\\
y\Eqn{=}i\Bigl(\Lambda^3-\Lambda^{-3}\Bigr) u^2
  +i\Bigl(3 \Lambda^6+\chi_1^{E_6} \Lambda^2
  -\chi_6^{E_6} \Lambda^{-2}-3 \Lambda^{-6}\Bigr) u\no\\
&&+i\left(2 \Lambda^9+2 \chi_1^{E_6} \Lambda^5
  -\chi_2^{E_6} \Lambda^3
  +(\chi_3^{E_6}+\chi_6^{E_6}) \Lambda\right.\no\\
&&\left.\hspace{2em}+(-\chi_1^{E_6}-\chi_5^{E_6})\Lambda^{-1}
  +\chi_2^{E_6} \Lambda^{-3}-2\chi_6^{E_6} \Lambda^{-5}
  -2 \Lambda^{-9}\right).
\end{eqnarray}
where $\chi_1^{E_6},\ldots,\chi_6^{E_6}$ denote
$E_6$ fundamental representation characters and are defined as
\ba
&&
\chi_1^{E_6}=\chEsi{1}{0}{0}{0}{0}{0}, \ \
\chi_2^{E_6}=\chEsi{0}{1}{0}{0}{0}{0}, \ \
\chi_3^{E_6}=\chEsi{0}{0}{1}{0}{0}{0}, \ \
\chi_4^{E_6}=\chEsi{0}{0}{0}{1}{0}{0},\no\\
&&
\chi_5^{E_6}=\chEsi{0}{0}{0}{0}{1}{0}, \ \
\chi_6^{E_6}=\chEsi{0}{0}{0}{0}{0}{1}.
\ea

\vspace{1ex}
\noindent
$126$ sections for $E_7$ curve:
\begin{eqnarray}
x\Eqn{=}\frac{1}{4\sin^2\lambda}\times\no\\
&&\Bigl[u^2+\Bigl(2 \chi_6^{D_6} \cos\lambda\Bigr) u
  +\Bigl(2 \cos 4\lambda+(2 \chi_5^{D_6}-20) \cos 2\lambda\no\\
&&\hspace{1.63em}
  +\chi_6^{D_6}\chi_6^{D_6}-2 \chi_5^{D_6}+18\Bigr)\Bigr],\ea
\ba
y\Eqn{=}\frac{1}{4\sin^3\lambda}\times\no\\
&&\Bigl[(\cos\lambda)u^3
  +\Bigl(\chi_6^{D_6}\cos 2\lambda+2 \chi_6^{D_6}\Bigr)u^2\no\\
&&\ 
  +\Bigl(-\cos 5\lambda+(\chi_5^{D_6}+5) \cos 3\lambda
  -2 \chi_2^{D_6} \cos 2\lambda\no\\
&&\hspace{1.63em}+(3 \chi_6^{D_6}\chi_6^{D_6}
  -\chi_5^{D_6}-4) \cos\lambda+2 \chi_2^{D_6}\Bigr) u\no\\
&&\ 
  +\Bigl(-\chi_1^{D_6} \cos 5\lambda+(\chi_4^{D_6}+\chi_6^{D_6}) \cos 4\lambda
  +(-\chi_2^{D_6} \chi_6^{D_6}+3 \chi_1^{D_6}) \cos 3\lambda\no\\
&&\hspace{1.63em}
  +(2 \chi_5^{D_6} \chi_6^{D_6}-4 \chi_4^{D_6}) \cos 2\lambda
  +(\chi_2^{D_6} \chi_6^{D_6}-2 \chi_1^{D_6}) \cos\lambda\no\\
&&\hspace{1.63em}
  +\chi_6^{D_6}\chi_6^{D_6}\chi_6^{D_6}-2 \chi_5^{D_6} \chi_6^{D_6}
  +3 \chi_4^{D_6}-\chi_6^{D_6}\Bigr)\Bigr]
\end{eqnarray}
where $\chi_1^{D_6},\ldots,\chi_6^{D_6}$ denote
$D_6$ fundamental representation characters and are defined as
\ba
&&
\chi_1^{D_6}=\chDsi{1}{0}{0}{0}{0}{0}, \ \
\chi_2^{D_6}=\chDsi{0}{1}{0}{0}{0}{0}, \ \
\chi_3^{D_6}=\chDsi{0}{0}{1}{0}{0}{0}, \ \
\chi_4^{D_6}=\chDsi{0}{0}{0}{1}{0}{0},\no\\
&&
\chi_5^{D_6}=\chDsi{0}{0}{0}{0}{1}{0}, \ \
\chi_6^{D_6}=\chDsi{0}{0}{0}{0}{0}{1}.
\ea

\section{Special $E_8$ Curves with Singular Fibers}

\noindent
$\bullet\ \vec{m}=\frac{2\pi}{2}\vec{\mu}_1
  =(0,0,0,0,0,0,0,2\pi)
\quad\Bigl(\mbox{or}\ \vec{m}=(0,0,0,0,\pi,\pi,\pi,\pi)\Bigr)$
\ba
f\Eqn{=}\frac{1}{12}(u-64)^2(u^2+128u-8192),\\
g\Eqn{=}\frac{1}{216}(u-64)^3(u-32)(u^2-640u+28672),\\[1ex]
\Delta\Eqn{=}(u-64)^{10}(u-48).
\ea
\quad$\Rightarrow$\quad degenerate fiber: $I_4^\ast$\qquad symmetry: $D_8$

\noindent
$\bullet\ \vec{m}=\frac{2\pi}{3}\vec{\mu}_2
  =(\frac{\pi}{3},\frac{\pi}{3},\frac{\pi}{3},\frac{\pi}{3},
    \frac{\pi}{3},\frac{\pi}{3},\frac{\pi}{3},\frac{5\pi}{3})
\quad\Bigl(\mbox{or}\ \vec{m}
  =(\frac{2\pi}{3},\frac{2\pi}{3},\frac{2\pi}{3},\frac{2\pi}{3},
    \frac{2\pi}{3},\frac{2\pi}{3},\frac{2\pi}{3},-\frac{2\pi}{3})\Bigr)$
\ba
f\Eqn{=}\frac{1}{12}(u-54)(u^3+54u^2-16524u+583200),\\
g\Eqn{=}\frac{1}{216}u^6-4u^5+873u^4-\frac{172179}{2}u^3+4376916u^2\no\\ \Eqn{}
  -112140612u+1147184289,\\
\Delta\Eqn{=}(u-63)^9(u^2-117u+3429).
\ea
\quad$\Rightarrow$\quad degenerate fiber: $I_9$\qquad symmetry: $A_8$

\noindent
$\bullet\ \vec{m}=\frac{2\pi}{4}\vec{\mu}_3
  =(-\frac{\pi}{4},\frac{\pi}{4},\frac{\pi}{4},\frac{\pi}{4},
    \frac{\pi}{4},\frac{\pi}{4},\frac{\pi}{4},\frac{7\pi}{4})
\quad\Bigl(\mbox{or}\ \vec{m}
  =(\frac{\pi}{4},\frac{\pi}{4},\frac{\pi}{4},\frac{\pi}{4},
    \frac{\pi}{4},\frac{\pi}{4},\frac{\pi}{4},-\frac{7\pi}{4})\Bigr)$
\ba
f\Eqn{=}\frac{1}{12}u^4-\frac{4864}{3}u^2+122880u-\frac{7847936}{3},\\
g\Eqn{=}\frac{1}{216}(u^2-96u+2176)(u^4-768u^3+112640u^2\no\\ \Eqn{} -6094848u+113115136),
\\
\Delta\Eqn{=}(u-64)^8(u-56)^2(u-60).
\ea
\quad$\Rightarrow$\quad degenerate fiber: $I_8,\ I_2$
\qquad symmetry: $A_7\oplus A_1$

\noindent
$\bullet\ \vec{m}=\frac{2\pi}{6}\vec{\mu}_4
  =(0,0,\frac{\pi}{3},\frac{\pi}{3},\frac{\pi}{3},\frac{\pi}{3},
    \frac{\pi}{3},\frac{5\pi}{3})$
\ba
f\Eqn{=}\frac{1}{12}(u-54)(u^3+54u^2-16308u+569160),\\
g\Eqn{=}\frac{1}{216}(u^2-96u+2196)(u^4-768u^3+112104u^2\no\\ \Eqn{}-6035904u+111493584),
\\[1ex]
\Delta\Eqn{=}(u-66)^6(u-57)^3(u-58)^2.
\ea
\quad$\Rightarrow$\quad degenerate fiber: $I_6,\ I_3,\ I_2$
\qquad symmetry: $A_5\oplus A_2\oplus A_1$

\noindent
$\bullet\ \vec{m}=\frac{2\pi}{5}\vec{\mu}_5
  =(0,0,0,\frac{2\pi}{5},\frac{2\pi}{5},\frac{2\pi}{5},
    \frac{2\pi}{5},\frac{8\pi}{5})$
\ba
f\Eqn{=}\frac{1}{12}u^4-\frac{4750}{3}u^2+118750u-\frac{7512500}{3},\\
g\Eqn{=}\frac{1}{216}(u^2-114u+3250)(u^4-750u^3+98750u^2\no\\ \Eqn{}-4687500u+74687500),
\\[1ex]
\Delta\Eqn{=}(u^2-125u+3875)^5(u-57).
\ea
\quad$\Rightarrow$\quad degenerate fiber: $I_5,\ I_5$
\qquad symmetry: $A_4\oplus A_4$

\noindent
$\bullet\ \vec{m}=\frac{2\pi}{4}\vec{\mu}_6
  =(0,0,0,0,\frac{\pi}{2},\frac{\pi}{2},
    \frac{\pi}{2},\frac{3\pi}{2})
\quad\Bigl(\mbox{or}\ \vec{m}=(0,0,0,0,0,\pi,\pi,\pi)\Bigr)$
\ba
f\Eqn{=}\frac{1}{12}(u-56)^2(u^2+112u-9152),\\
g\Eqn{=}\frac{1}{216}(u-56)^3(u-40)(u^2-656u+33856),\\[1ex]
\Delta\Eqn{=}(u-56)^7(u-72)^4.
\ea
\quad$\Rightarrow$\quad degenerate fiber: $I_1^\ast,\ I_4$
\qquad symmetry: $D_5\oplus A_3$

\vspace{3ex}
\noindent
$\bullet\ \vec{m}=\frac{2\pi}{3}\vec{\mu}_7
  =(0,0,0,0,0,\frac{2\pi}{3},\frac{2\pi}{3},\frac{4\pi}{3})
\quad\Bigl(\mbox{or}\ \vec{m}
  =(0,0,0,0,0,\frac{4\pi}{3},\frac{4\pi}{3},\frac{4\pi}{3})\Bigr)$
\ba
f\Eqn{=}\frac{1}{12}(u-54)^3(u+162),\\
g\Eqn{=}\frac{1}{216}(u-54)^4(u^2-648u+26244),\\[1ex]
\Delta\Eqn{=}(u-54)^8(u-81)^3.
\ea
\quad$\Rightarrow$\quad degenerate fiber: $IV^\ast,\ I_3$
\qquad symmetry: $E_6\oplus A_2$

\noindent
$\bullet\ \vec{m}=\frac{2\pi}{2}\vec{\mu}_8
  =(0,0,0,0,0,0,\pi,\pi)
\quad\Bigl(\mbox{or}\ \vec{m}
  =(\frac{\pi}{2},\frac{\pi}{2},\frac{\pi}{2},\frac{\pi}{2},
    \frac{\pi}{2},\frac{\pi}{2},\frac{\pi}{2},\frac{\pi}{2})\Bigr)$
\ba
f\Eqn{=}\frac{1}{12}(u-48)^3(u+144),\\
g\Eqn{=}\frac{1}{216}(u-48)^5(u-624),\\[1ex]
\Delta\Eqn{=}(u-48)^9(u-112)^2.
\ea
\quad$\Rightarrow$\quad degenerate fiber: $III^\ast,\ I_2$
\qquad symmetry: $E_7\oplus A_1$
\noindent
$\vec{\mu}_1,\ldots,\vec{\mu}_8$ denotes the fundamental weights of
$E_8$:
\ba
\vec{\mu}_1\Eqn{=}2\vec{\bf e}_8,\\
\vec{\mu}_2\Eqn{=}\tfrac{1}{2}\vec{\bf e}_1+\tfrac{1}{2}\vec{\bf e}_2
  +\tfrac{1}{2}\vec{\bf e}_3+\tfrac{1}{2}\vec{\bf e}_4
  +\tfrac{1}{2}\vec{\bf e}_5+\tfrac{1}{2}\vec{\bf e}_6
  +\tfrac{1}{2}\vec{\bf e}_7+\tfrac{5}{2}\vec{\bf e}_8,\\
\vec{\mu}_3\Eqn{=}-\tfrac{1}{2}\vec{\bf e}_1+\tfrac{1}{2}\vec{\bf e}_2
  +\tfrac{1}{2}
\vec{\bf e}_3+\tfrac{1}{2}\vec{\bf e}_4+\tfrac{1}{2}\vec{\bf e}_5
  +\tfrac{1}{2}\vec{\bf e}_6+\tfrac{1}{2}\vec{\bf e}_7
  +\tfrac{7}{2}\vec{\bf e}_8,\\
\vec{\mu}_4\Eqn{=}\vec{\bf e}_3+\vec{\bf e}_4+\vec{\bf e}_5
  +\vec{\bf e}_6+\vec{\bf e}_7+5\vec{\bf e}_8,\\
\vec{\mu}_5\Eqn{=}\vec{\bf e}_4+\vec{\bf e}_5+\vec{\bf e}_6
  +\vec{\bf e}_7+4\vec{\bf e}_8,\\
\vec{\mu}_6\Eqn{=}\vec{\bf e}_5+\vec{\bf e}_6+\vec{\bf e}_7+3\vec{\bf e}_8,\\
\vec{\mu}_7\Eqn{=}\vec{\bf e}_6+\vec{\bf e}_7+2\vec{\bf e}_8,\\
\vec{\mu}_8\Eqn{=}\vec{\bf e}_7+\vec{\bf e}_8.
\ea
We show the labeling of the fundamental weights of $E_8$:
\begin{figure}[h]
\begin{center}
\unitlength=2.4pt
\begin{picture}(60,30)
\put( 1,10){\line(1,0){8}}
\put(20,11){\line(0,1){8}}
\put(11,10){\line(1,0){8}}
\put(21,10){\line(1,0){8}}
\put(31,10){\line(1,0){8}}
\put(41,10){\line(1,0){8}}
\put(51,10){\line(1,0){8}}
\put( 0,10){\circle{2}}
\put(20,20){\circle{2}}
\put(10,10){\circle{2}}
\put(20,10){\circle{2}}
\put(30,10){\circle{2}}
\put(40,10){\circle{2}}
\put(50,10){\circle{2}}
\put(60,10){\circle{2}}
\put( 0, 6){${\!\!\!\!\>}_{(1)}$}
\put(20,20){${\hspace{0.4em}}_{(2)}$}
\put(10, 6){${\!\!\!\!\>}_{(3)}$}
\put(20, 6){${\!\!\!\!\>}_{(4)}$}
\put(30, 6){${\!\!\!\!\>}_{(5)}$}
\put(40, 6){${\!\!\!\!\>}_{(6)}$}
\put(50, 6){${\!\!\!\!\>}_{(7)}$}
\put(60, 6){${\!\!\!\!\>}_{(8)}$}
\end{picture}
\end{center}
\vspace{-3ex}
\caption{Dynkin diagram for $E_8$}
\end{figure}
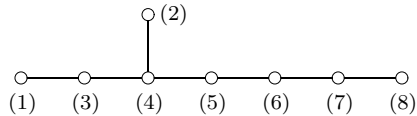

\renewcommand{\section}{\subsection}
\renewcommand{\refname}

\end{document}